\documentclass[12pt]{article}
\pdfoutput=1

\usepackage[a4paper,text={16.8cm,22.4cm}]{geometry}
\usepackage{amsmath,amsfonts,slashed,amssymb,tikz,bm,psfrag,graphicx,color,dsfont}
\usepackage{multicol,multirow}
\usepackage{float}

\RequirePackage[sort&compress,square,comma,numbers]{natbib}
\allowdisplaybreaks
\renewcommand{\arraystretch}{1.2}

\begin{document}

\begin{titlepage}

\begin{flushright}
\normalsize
February 10, 2020
\end{flushright}

\vspace{0.1cm}
\begin{center}
\Large\bf
QCD calculations of radiative heavy meson decays
with subleading power corrections
\end{center}

\vspace{0.5cm}
\begin{center}
{\bf Hua-Dong Li$^{a}$,
Cai-Dian L\"{u}$^{a,b}$,
Chao Wang$^{c}$,
Yu-Ming Wang$^{c}$£¬
Yan-Bing Wei$^{c}$}\\
\vspace{0.7cm}
{\sl   ${}^a$ \, Institute of High Energy Physics, CAS, P.O. Box 918(4) Beijing 100049,  China \\
 ${}^b$ \,  School of Physics, University of Chinese Academy of Sciences, Beijing 100049,  China \\
${}^c$ School of Physics, Nankai University, Weijin Road 94, 300071 Tianjin, China \,
}
\end{center}

\vspace{0.2cm}
\begin{abstract}

We revisit QCD calculations of radiative heavy meson decay form factors by including the subleading
power corrections from the twist-two photon distribution amplitude at next-to-leading-order
in $\alpha_s$ with the method of the light-cone sum rules (LCSR).
The desired hard-collinear factorization formula for the vacuum-to-photon correlation function
with the interpolating currents for two heavy mesons is constructed with the operator-product-expansion technique
in the presence of evanescent operators.
Applying the background  field approach, the higher twist corrections from both the two-particle
and three-particle photon distribution amplitudes are further computed in the LCSR framework at leading-order in QCD,
up to the twist-four accuracy.
Combining the leading power ``point-like" photon contribution at tree level
and the subleading power resolved photon corrections from the newly derived  LCSR,
we update theory predictions for the nonperturbative couplings describing the electromagnetic
decay processes of the heavy mesons $H^{\ast \, \pm} \to H^{\pm} \, \gamma$, $H^{\ast \, 0} \to H^{0} \, \gamma$,
$H_s^{\ast \, \pm} \to H_s^{\pm} \, \gamma$ (with $H=D, \,  B$).
Furthermore, we perform an exploratory comparisons of our sum rule computations of the heavy-meson magnetic couplings
with the previous determinations  based upon  different QCD approaches and phenomenological models.

\end{abstract}

\vfil

\end{titlepage}

\tableofcontents

\newpage

\section{Introduction}

Advancing our understanding of radiative heavy  meson  decays is, on the one hand,  of importance for exploring the
emerged symmetries of the QCD Lagrangian in the limit of massless light quarks
and of infinitely heavy quarks, on the other hand,  crucial to develop a systematic formalism for
computing the electromagnetic corrections to flavour-changing weak decays of heavy hadrons,
which are indispensable for a detailed anatomy of the quark flavour dynamics of the Standard Model (SM).
On the phenomenological aspects, precision calculations of radiative heavy  meson  decays
are also in high demand for the sake of determining the magnetic susceptibility of the quark condensate
$\chi(\mu)$ \cite{Ioffe:1983ju}, which parameterizes the response of the QCD vacuum with respect to an external electromagnetic field
and serves as an essential nonperturbative input for the theory description of exclusive hadronic reactions involving on-shell photons
\cite{Balitsky:1989ry,He:2006ud,Wang:2008sm,Mannel:2011xg,Wang:2015ndk,Ball:2003fq,
Wang:2018wfj,Khodjamirian:2010vf,Li:2013xna,Wang:2017ijn,Li:2010nn}
and for the improved calculations of the muon anomalous magnetic moment \cite{Czarnecki:2002nt}.
Consequently, distinct QCD techniques have been developed to allow for the systematical computations of
radiative heavy  meson  decay amplitudes based upon the heavy quark expansion technique and perturbative factorization theorems.

Employing the method of the two-point QCD sum rules (QCDSR),
the electromagnetic  $D^{\ast}(p+q) \to D(q) \, \gamma(p)$  decay form factors
were estimated at leading-order (LO) in $\alpha_s$ \cite{Eletsky:1984qs}
by taking into account the subleading power corrections at the dimension-5
quark-gluon condensate accuracy. However,  implementing an approximation  to avoid the
double Borel transformation with respect to the variables $p^2$ and $(p+q)^{2}$
\cite{Eletsky:1984qs} yields contamination of the obtained sum rules due to the ``non-diagonal"
transitions of the ground state to  excited states.
The resulting prediction for the branching fraction of $D^{\ast 0} \to D^{0} \, \gamma$ turned out to be
larger than that for the strong decay process $D^{\ast 0} \to D^{0} \, \pi^{0}$ \cite{Eletsky:1984qs},
in contradiction with the experimental measurements from the CLEO \cite{Butler:1992rp,Bartelt:1997yu},
BaBar \cite{Aubert:2005ik}, and BES III \cite{Ablikim:2014mww} Collaborations.
Subsequently, the radiative charm-meson decay form factors were computed from the three-point QCDSR approach
\footnote{The double dispersion sum rules for computing radiative transition form factors in QCD 
were originally suggested in \cite{Khodjamirian:1979fa}.},
including the power suppressed corrections from the higher-dimension operators up to the four-quark condensate \cite{Aliev:1994nq},
yielding the theoretical predictions in reasonable agreement with the experimental results
\cite{Butler:1992rp,Bartelt:1997yu,Aubert:2005ik,Ablikim:2014mww}.
As demonstrated in \cite{Aliev:1994nq}, the numerically dominant contribution to the  tree-level sum rules for the radiative
$D_{(s)}^{\ast +} \to D_{(s)}^{+} \, \gamma$ form factors  arises from the dimension-3 quark condensate correction
instead of the leading power perturbative effect, due to the strong cancellation for the photon radiation off
the charm and the anti-down (anti-strange) quarks, which also justifies the high suppression of the
${D_{(s)}^{\ast +} D_{(s)}^{+} \, \gamma}$ coupling compared with the magnetic moment for
the counterpart neutral $D^{\ast 0}$-meson.
In an attempt to eliminate the substantial contamination from the non-diagonal transitions
of constructing the traditional QCDSR  for hadronic matrix elements at small momentum transfer,
the technically improved sum rules based upon the light-cone operator-product-expansion (OPE) for
the corresponding QCD correlation functions have been constructed \cite{Aliev:1995zlh} for computing the radiative
heavy meson decay form factors with the subleading power corrections from the photon light-cone distribution
amplitudes (LCDA) at the (partial)-twist-four accuracy.
Motivated by the systematic classification of the two-particle and three-particle photon distribution amplitudes
with the background field formalism \cite{Ball:2002ps},
the updated light-cone sum rules (LCSR) for the electromagnetic heavy-meson decay form factors
were further derived in \cite{Rohrwild:2007yt} including the two-particle and
three-particle ``hadronic" photon corrections at twist-four completely.

Taking advantage of the spontaneously broken chiral symmetry for the light quarks and
the emerged spin-flavour symmetry for the heavy quarks,
systematic computations of the radiative heavy-meson decay form factors
have been also carried out in the framework of
heavy hadron chiral perturbation theory (HH$\chi$PT) \cite{Amundson:1992yp,Cho:1992nt,Cheng:1992xi},
including the one-loop corrections to symmetry breaking effects
at ${\cal O}(m_q)$  and ${\cal O}(1/m_Q)$ \cite{Stewart:1998ke}
(see \cite{Casalbuoni:1996pg,Manohar:2000dt} for further discussions).
The striking large SU(3)-flavour symmetry breaking effects for the $D_q^{\ast} \to D_q \, \gamma$ (with $q=u, d, s$)
decay rates can be apparently  traced back to the numerical cancellation between the leading power and
the subleading  power contributions in  the heavy quark expansion, by accident,  for the charged charm-meson form factors
with  physical values of  the charm-quark mass.  However, the two effective parameters $g_{\pi}$ and $\beta$
characterizing the $M^{\ast}   M \, \pi$ and  $M^{\ast}  M \, \gamma$ couplings (with $M=D_{(s)}, \, B_{(s)}$)
can only be extracted from the available experimental measurements of the corresponding vector charm-meson decays
\cite{Stewart:1998ke} or determined by the non-perturbative QCD techniques
(see \cite{ElBennich:2010ha,Bernardoni:2014kla,Becirevic:2009xp,Donald:2013sra,Colangelo:2000dp} and references therein),
generating a significant limitation of the predictive power for the HH$\chi$PT formalism.
We further  mention in passing that the electromagnetic transition form factors of the low-lying heavy mesons
have been evaluated phenomenologically  \cite{Colangelo:1993zq} by incorporating the HH$\chi$PT  framework
and the vector meson dominance (VMD) hypothesis, which gives rise to the potentially sizeable systematic uncertainty
of the resulting predictions and should be improved upon by employing the dispersion approach
as developed in \cite{Khodjamirian:1997tk}  (see \cite{Khodjamirian:2010vf,Khodjamirian:2012rm,Wang:2016qii} for more applications).

On account of the substantial power suppressed corrections to the radiative  $D_{(s)}^{\ast +} \to D_{(s)}^{+} \, \gamma$ decay
form factors, it is apparently of importance to  compute  perturbative QCD corrections to the resolved photon contributions
to  all the magnetic couplings responsible for $H^{\ast \, \pm} \to H^{\pm} \, \gamma$, $H^{\ast \, 0} \to H^{0} \, \gamma$,
$H_s^{\ast \, \pm} \to H_s^{\pm} \, \gamma$ (with $H=D, \,  B$) at the  twist-two  accuracy
and to refine the aforementioned higher twist hadronic photon corrections at tree level
presented in \cite{Aliev:1995zlh,Rohrwild:2007yt}  by employing the LCSR method,
in order to achieve a better understanding of the heavy quark expansion for the charm and bottom hadron decays
and to provide the state-of-art theory predictions for the heavy-meson  electromagnetic transition form factors in QCD.
We summarize the essential new ingredients of the present paper as follows.

\begin{itemize}

\item{We establish the hard-collinear factorization formula for the twist-two contribution to the
vacuum-to-photon correlation function  defined by the vector and pseudoscalar interpolating currents for two heavy mesons
at next-to-leading-order (NLO) in QCD, with the aid of the evanescent operator approach \cite{Dugan:1990df,Herrlich:1994kh}
and the strategy of regions \cite{Beneke:1997zp,Smirnov:2002pj}.
The double spectral density appearing in the dispersion representation of the obtained  factorization formula
will be derived analytically for the asymptotic photon distribution amplitude $\phi_{\gamma}^{\rm asy}(u, \mu) = 6 \, u \, (1-u)$,
following the prescriptions introduced in \cite{Khodjamirian:1999hb}.
In particular, we extract the general expression for the QCD spectral function of the NLO correction
to the  twist-two hadronic photon contribution of the vacuum-to-photon correlation function,
without implementing the further reduction dependent on the specific shape of the duality region
for the resulting dispersion integral \cite{Belyaev:1994zk}.
QCD resummation for the enhanced logarithms of $m_Q/\Lambda_{\rm QCD}$ entering the hard-collinear factorization formula
will be accomplished at next-to-leading-logarithmic (NLL) accuracy by solving the renormalization-group (RG)
equation of $\phi_{\gamma}(u, \mu)$ at two loops.}

\item{We perform the continuum subtraction  for constructing the sum rules of the higher twist corrections
to the  $M^{\ast} M \gamma$ couplings by computing the corresponding double spectral densities at LO in $\alpha_s$
analytically, instead of invoking the simple replacement rule  valid  for the leading twist contribution
as employed in \cite{Aliev:1995zlh,Rohrwild:2007yt}.
Our method can be also generalized to the LCSR calculation of the $M^{\ast} M \pi$ couplings straightforwardly. }

\item{We compute the light-quark mass effect for  the ``point-like" photon contribution
to the radiative heavy meson  decay form factors at tree level and demonstrate explicitly
that such SU(3)-flavour symmetry breaking effect is power suppressed in the heavy quark expansion,
in contrast to the observed pattern for heavy-to-light $B$-meson decay form factors
at large hadronic recoil \cite{Lu:2018cfc,Gao:2019lta}.
Furthermore, we identify the precise correspondence for all separate terms entering
the ``point-like" photon contributions  to the  $M^{\ast} M \gamma$ couplings
between the LCSR calculation and the HH$\chi$PT analysis.}

\end{itemize}

This paper is structured as follows. In Section \ref{section: theory summary}
we will briefly  review the  two equivalent definitions of the heavy meson  magnetic couplings
responsible for the radiative $M^{\ast} \to M \, \gamma$ transitions,
and then summarize the ``point-like"  photon contributions to the tree-level sum rules
of  the $M^{\ast}  M \, \gamma$ couplings by exploring the factorization properties of
the  vacuum-to-photon correlation functions under discussion in detail and  computing
the subleading power perturbative correction due to the non-vanishing light quark mass
at LO in the strong coupling constant $\alpha_s$.
We will proceed to establish the hard-collinear factorization formula for the  twist-two
resolved photon contribution to the vacuum-to-photon correlation function at one loop
with the OPE technique in Section \ref{section: leading-twist hadronic photon effect},
where the general expression of the double spectral density  for the obtained QCD representation
of the above-mentioned correlation function at NLO will be  derived analytically and the resummation
improved LCSR for the twist-two hadronic photon correction will be achieved with the standard RG resummation
technique in momentum space.
Employing the background field approach we will turn to compute the subleading power contributions to
the radiative heavy meson decay form factors from the two-particle and three-particle higher twist photon
distribution amplitudes at the twist-four accuracy within the same LCSR framework
in Section \ref{section: higher-twist hadronic photon correction}, with an emphasis on the technical
derivation of the  perturbative double spectral functions in the corresponding dispersion integrals.
Phenomenological implications of the refined LCSR for the $M^{\ast}  M \, \gamma$ couplings
will be subsequently explored in Section \ref{section: numerical analysis}, by investigating the numerical impacts
of the distinct subleading power contributions with  seven non-perturbative parameters
appearing in the conformal expansion of the collinear photon LCDA determined
from the QCDSR method \cite{Ball:2002ps,Balitsky:1989ry} and by performing an exploratory comparison
of our LCSR calculations with the previous determinations from the HH$\chi$PT method and from the lattice QCD simulation.
Section \ref{section: conclusion} will be reserved for a summary of our main observations
on the resolved photon corrections and for theory perspectives on the future improvements
of QCD calculations of the electromagnetic heavy meson transition form factors.
We further collect the useful results of various one-loop Feynman integrals and
the  master formulae to derive the spectral representation of the twist-two factorization formulae
for  the vacuum-to-photon correlation function at NLO
in Appendices \ref{appendix: Feynman integrals at one loop} and
\ref{appendix: master formulae for double spectral densities},  respectively.
For completeness,  we also summarize the definitions of the two-particle and three-particle
collinear photon LCDA at twist-four 
in Appendix \ref{appendix: Photon distribution amplitudes}.

\section{Theory summary for radiative heavy meson decays}
\label{section: theory summary}

The purpose of this section is to present the tree-level LCSR for  the ``point-like" photon contribution to
the $M^{\ast}  M \, \gamma$ coupling including the SU(3)-flavour symmetry breaking effect
due to the light-quark mass correction. The  vector-to-pseudoscalar heavy meson electromagnetic
transition matrix element will be parameterized in the standard way
\begin{eqnarray}
\langle M(q) |j_{\mu}^{\rm em} |  M^{\ast}(p+q, \varepsilon) \rangle
= g_{\rm em} \, {2 \, i \, {\cal V}(p^2) \over m_V + m_P} \, \epsilon_{\mu \nu \rho \sigma} \,
\varepsilon^{\nu} \, p^{\rho} \,  q^{\sigma} \,,
\end{eqnarray}
where we have introduced the following conventions
\begin{eqnarray}
j_{\mu}^{\rm em} =  g_{\rm em} \, \sum_{q} e_q \, \bar q \, \gamma_{\mu} \, q \,, \qquad
\epsilon_{0123} = - 1 \,,
\end{eqnarray}
and $m_{V(P)}$ is the heavy vector (pseudoscalar) meson mass.
The on-shell photon coupling with the vector and pseudoscalar heavy mesons defined as
\begin{eqnarray}
\langle \gamma(p, \eta^{\ast}) M(q)|M^{\ast}(p+q) \rangle
= - g_{\rm em} \, g_{M^{\ast} M \gamma} \, \epsilon_{\mu \nu \rho \sigma} \, \eta^{\ast \mu} \, \varepsilon^{\nu} \,
\, p^{\rho} \,  q^{\sigma}  \,
\label{definition: strong coupling}
\end{eqnarray}
can be readily deduced from the aforementioned  electromagnetic decay form factor
\begin{eqnarray}
g_{M^{\ast} M \gamma} =  { 2 \over m_V + m_P} \,  {\cal V} (p^2=0).
\end{eqnarray}

\begin{figure}
\begin{center}
\includegraphics[width=0.95 \columnwidth]{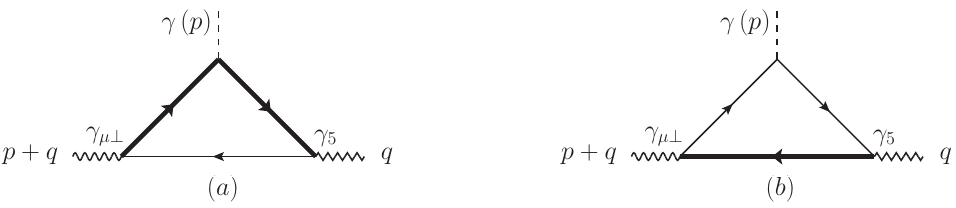}
\vspace*{0.1cm}
\caption{Diagrammatic representation of the ``point-like"  photon contribution to the
vacuum-to-photon correlation function (\ref{defnition: correlation function}) at tree level.}
\label{fig: LP effect}
\end{center}
\end{figure}

The general strategy of constructing the sum rules for the coupling $g_{M^{\ast} M \gamma}$
is to explore the vacuum-to-photon correlation function, following closely \cite{Wang:2018wfj},
defined with the two local interpolating currents for the vector and pseudoscalar heavy mesons
\begin{eqnarray}
\Pi_{\mu}(p, q) = \int d^4 x \, e^{-i (p+q) \cdot x} \,
\langle \gamma(p, \eta^{\ast})  |  {\rm T }  \left \{  \bar q(x) \gamma_{\mu \perp} Q(x),
\bar Q(0)  \gamma_5 q(0) \right \}  | 0 \rangle  \,,
\label{defnition: correlation function}
\end{eqnarray}
where the perpendicular components of the Dirac $\gamma$ matrices are given by
\begin{eqnarray}
\gamma_{\mu \perp} = \gamma_{\mu} - {\not \! n \over 2}\, \bar n_{\mu}
- {\not \! \bar n \over 2}\,  n_{\mu}, \qquad p_{\mu}= {n \cdot p \over 2} \, \bar n_{\mu} \,,
\end{eqnarray}
with the light-cone vectors $n_{\mu}$ and $\bar n_{\mu}$ satisfying the constraints
\begin{eqnarray}
n \cdot \bar n =2, \qquad n^2 = \bar n^2 =0 \,.
\end{eqnarray}To demonstrate the hard-collinear factorization formula for
the correlation function (\ref{defnition: correlation function}) we will employ
the following power counting scheme for the external momenta
\begin{eqnarray}
n \cdot p \sim {\cal O}(m_Q), \qquad |(p+q)^2 - m_Q^2| \sim  {\cal O}(m_Q^2),
\qquad |q^2 - m_Q^2| \sim  {\cal O}(m_Q^2) \,.
\label{power counting scheme}
\end{eqnarray}
The leading power perturbative contribution to the vacuum-to-photon correlation function
stems from the ``point-like"  photon component corresponding to the twist-one
distribution amplitude \cite{Ball:2002ps} and can be further derived at tree level straightforwardly
by evaluating  the two triangle diagrams corresponding to the collinear photon coupling with  the heavy-quark
and light-quark electromagnetic currents as displayed in figure \ref{fig: LP effect}.

The yielding contribution from the perturbative diagram in figure \ref{fig: LP effect}(a) is given by
\begin{eqnarray}
\Pi_{\mu}^{\rm 1(a)}(p, q) &=& - N_c \, e_{Q}  \, g_{\rm em} \, \int {d^D \ell \over (2\, \pi)^D} \,
\frac{1}{[(\ell - p -q)^2 + i 0 ] [\ell^2 - m_Q^2 + i 0] [(\ell-p)^2 - m_Q^2 + i 0]} \nonumber \\
&& {\rm Tr} \left [ \gamma_{\mu \perp} \, (\not \! \ell \, - \not \! p \, - \not \! q + m_q) \,
\gamma_5 \, (\not \! \ell \, - \not \! p + m_Q) \, \not \! \eta^{\ast} \, (\not \! \ell + m_Q) \right ]   \,,
\label{expression of the diagram 1a}
\end{eqnarray}
including the subleading power correction due to the light-quark mass at  ${\cal O}(\Lambda_{\rm QCD}/m_Q)$.
According to the power counting scheme (\ref{power counting scheme}) we can identify the leading power
contribution of the following scalar loop integral
\begin{eqnarray}
I_1 =  \int {d^D \ell \over (2\, \pi)^D} \,
\frac{1}{[(\ell - p -q)^2 + i 0 ] [\ell^2 - m_Q^2 + i 0] [(\ell-p)^2 - m_Q^2 + i 0]}
\end{eqnarray}
from the hard region in agreement with our expectation.
It is then evident that the loop-momentum integration entering (\ref{expression of the diagram 1a})
is free of the ultraviolet and infrared divergences. Computing  the Dirac algebra in
the four-dimensional space leads to
\begin{eqnarray}
\Pi_{\mu}^{\rm 1(a)}(p, q) &=& - 4 \, i \, N_c \, e_{Q}  \, g_{\rm em} \, m_Q \, \int {d^D \ell \over (2\, \pi)^D} \,
\frac{1}{[(\ell - p -q)^2 + i 0 ] [\ell^2 - m_Q^2 + i 0] [(\ell-p)^2 - m_Q^2 + i 0]}
\nonumber \\
&& \times \, \left [ 1 - (1-r_q) \, {\ell \cdot p  \over p \cdot q}   \right ] \,
\epsilon_{\mu \, p \,  q \, \eta} \,,
\label{reduced expression of the diagram 1a}
\end{eqnarray}
with $r_q=m_q/m_Q$. Implementing the loop momentum integration subsequently  yields
\begin{eqnarray}
\Pi_{\mu}^{\rm 1(a)}(p, q) &=&  - \left ( {N_c \over 4 \, \pi^2} \right ) \, e_Q \, g_{\rm em} \, m_Q \,
\epsilon_{\mu \, p \,  q \, \eta} \, \int_0^1 dy \, \int_0^1 dt \,
{\bar y + y \, r_q \over m_Q^2 - y \, \bar t \, (p+q^2) - y \, t \, q^2}\,,
\label{result of the diagram 1a}
\end{eqnarray}
where we have introduced the notations $\bar y = 1-y, \, \bar t = 1- t$ for brevity.

Along the same vein we can write down the perturbative contribution to the correlation
function  (\ref{defnition: correlation function}) due to the real photon radiation off the
light quark
\begin{eqnarray}
\Pi_{\mu}^{\rm 1(b)}(p, q) &=& - N_c \, e_{q}  \, g_{\rm em} \, \int {d^D \ell \over (2\, \pi)^D} \,
\frac{1}{[(\ell - p -q)^2 - m_Q^2 + i 0 ] [\ell^2  + i 0] [(\ell-p)^2 + i 0]} \nonumber \\
&& {\rm Tr} \left [ \gamma_{\mu \perp} \, (\not \! \ell \, - \not \! p \, - \not \! q + m_Q) \,
\gamma_5 \, (\not \! \ell \, - \not \! p + m_q) \, \not \! \eta^{\ast} \, (\not \! \ell + m_q) \right ]   \,, \nonumber \\
&=& N_c \, e_{q}  \, g_{\rm em} \, \int {d^D \ell \over (2\, \pi)^D} \,
\frac{1}{[(\ell - p -q)^2 - m_Q^2 + i 0 ] [\ell^2  + i 0] [(\ell-p)^2 + i 0]} \nonumber \\
&& \left  \{ m_Q \, {\rm Tr} \left [ \gamma_{\mu \perp} \, \gamma_5 \,
\not \! p  \, \not \! \eta^{\ast} \, \not \! \ell \right ]
- m_q \,  {\rm Tr} \left [ \gamma_{\mu \perp} \,
\gamma_5 \, \not \! p  \, \not \! \eta^{\ast} \, (\not \! \ell \, - \not \! q)  \right ] \right \}  \,,
\label{expression of the diagram 1b}
\end{eqnarray}
where we have  explicitly  separated the heavy-quark mass term from the light-quark mass correction
to facilitate the power counting analysis for the diagram presented in figure \ref{fig: LP effect}(b).
We are now in a position to determine all regions of the loop momentum yielding the leading power contributions
to the two Feynman integrals  in (\ref{expression of the diagram 1b})
\begin{eqnarray}
I_{2, A} &=&  \int {d^D \ell \over (2\, \pi)^D} \,
\frac{\bar n \cdot \ell}{[(\ell - p -q)^2 - m_Q^2 + i 0 ] [\ell^2 + i 0] [(\ell-p)^2  + i 0]} \,, \\
I_{2, B} &=&  \int {d^D \ell \over (2\, \pi)^D} \,
\frac{\bar n \cdot (\ell-q)}{[(\ell - p -q)^2 - m_Q^2 + i 0 ] [\ell^2 + i 0] [(\ell-p)^2  + i 0]} \,.
\end{eqnarray}
Applying the default power counting scheme (\ref{power counting scheme}) it is straightforward to
verify that $I_{2, A}$ receives the non-vanishing contribution only from the hard loop-momentum region and
it therefore requires no ultraviolet and infrared subtractions  in analogy to the scalar loop integral $I_1$.
As a consequence, perturbative QCD factorization for the corresponding contribution to
the correlation function (\ref{defnition: correlation function}) can  indeed be established
at leading power in $\Lambda_{\rm QCD}/m_Q$.
By contrast, the remaining  Feynman integral $I_{2, B}$ due to the light quark mass correction can be contributed
from both the hard and collinear loop-momentum regions, where the typical scaling for a collinear momentum vector
$\ell_{\mu}$ reads
\begin{eqnarray}
\ell_{\mu} = (n \cdot \ell, \,\, \bar n \cdot \ell, \,\, \ell_{\perp \mu})
\sim (1, \,\, \lambda^2,  \,\, \lambda) \, m_Q, \qquad
\lambda = \Lambda_{\rm QCD}/m_Q\,.
\end{eqnarray}
The emergence of the leading contribution to $I_{2, B}$ from the collinear momentum region
implies that the nonperturbative  photon distribution amplitudes must be introduced in the factorization
formula to subtract the long-distance strong interaction effect from the perturbative QCD calculation
of the correlation function (\ref{defnition: correlation function}).
Since the SU(3)-flavour symmetry breaking effect for the obtained light-cone  matrix element
$\langle \gamma(p, \eta^{\ast})  | \bar q(z) \, \gamma_{\mu \perp}  \, \gamma_5 \,  q(0)|  0 \rangle$
from the infrared subtraction program (see \cite{Ball:1998ff,Ball:2007zt} for discussions
in the context of the vector meson distribution amplitudes)
has not been investigated systematically at present,
we will not take into account the light quark mass correction to the perturbative contribution
shown in figure \ref{fig: LP effect}(b) further,
which  is apparently suppressed by one power of $\Lambda_{\rm QCD}/m_Q$ in the heavy quark expansion.
Performing the loop momentum integration for the QCD amplitude $\Pi_{\mu}^{\rm 1(b)}$
explicitly gives rise to
\begin{eqnarray}
\Pi_{\mu}^{\rm 1(b)}(p, q) &\supset&  - \left ( {N_c \over 4 \, \pi^2} \right ) \, e_q \, g_{\rm em} \, m_Q \,
\epsilon_{\mu \, p \,  q \, \eta} \, \int_0^1 dy \, \int_0^1 dt \,
{y  \over m_Q^2 - y \, \bar t \, (p+q^2) - y \, t \, q^2} \,.
\label{result of the diagram 1b}
\end{eqnarray}

Adding up the two different pieces of perturbative contributions displayed in
(\ref{result of the diagram 1a}) and (\ref{result of the diagram 1b}) yields
the tree-level factorization formula for the ``point-like" photon contribution
\begin{eqnarray}
\Pi_{\mu}^{(\rm per)}(p, q) &\supset&
 - \left ( {N_c \over 4 \, \pi^2} \right )  \, g_{\rm em} \, m_Q \,
\epsilon_{\mu \, p \,  q \, \eta} \,
 \int_0^1 dy \, \int_0^1 dt \,
{e_Q \, \left ( \bar y + y \, r_q  \right ) + e_q \, y  \over m_Q^2 - y \, \bar t \, (p+q^2) - y \, t \, q^2} \,,
\label{QCD result of the point-like photon effect}
\end{eqnarray}
which  can be expressed in terms of the following dispersion integral
\begin{eqnarray}
\Pi_{\mu}^{(\rm per)}(p, q) &\supset&
 - \left ( {N_c \over 4 \, \pi^2} \right )  \, g_{\rm em} \, m_Q \,
\epsilon_{\mu \, p \,  q \, \eta} \,
 \int d s_1  \, \int d s_2  \, \frac{\rho^{(\rm per)}(s_1, s_2)}{[s_1 -(p+q)^2 - i 0] [s_2 - q^2 - i 0]} \,,
\label{QCD result of the point-like photon effect: dispersion form}
\end{eqnarray}
with the QCD spectral density at LO given by
\begin{eqnarray}
\rho^{(\rm per)}(s_1, s_2) &=& - \delta(s_1 - s_2) \,
\left \{e_Q \, \left [ (1-r_q) \, \left ( 1- {m_Q^2 \over s_2} \right )
+ \ln \left ( {m_Q^2 \over s_2} \right )  \right ]
- e_q \,  \left ( 1- {m_Q^2 \over s_2} \right )  \right \} \nonumber \\
&&  \times \, \theta(s_2 - m_Q^2) + {\cal O}(\alpha_s) \,.
\end{eqnarray}
Taking advantage of the standard definitions for the heavy meson decay constants
\begin{eqnarray}
\langle 0 | \bar Q \, \gamma_5 \, q | M(q) \rangle = - i \, {f_P \, m_P^2 \over m_Q + m_q}\,,
\qquad
\langle M^{\ast}(p+q) | \bar q \, \gamma_{\mu \perp} \, Q | 0 \rangle
= - i \, f_{V} \, m_V \, \varepsilon_{\mu \perp}^{\ast}
\end{eqnarray}
and for the coupling $g_{M^{\ast} M \gamma}$ as introduced in (\ref{definition: strong coupling}),
we can readily derive the hadronic dispersion relation for
the vacuum-to-photon correlation function (\ref{defnition: correlation function})
\begin{eqnarray}
\Pi_{\mu}(p, q) &=& - \epsilon_{\mu \, p \,  q \, \eta} \,
\bigg \{ \frac{g_{\rm em} \, f_P \, f_V \, m_V \, g_{M^{\ast} M \gamma}}{[m_V^2 -(p+q)^2 - i 0] [m_P^2 - q^2 - i 0]}
\, {m_P^2 \over m_Q + m_q} \,  \nonumber \\
&& \hspace{1.5 cm} +  \int d s_1  \, \int d s_2  \,
\frac{\rho^{(\rm had)}(s_1, s_2)}{[s_1 -(p+q)^2 - i 0] [s_2 - q^2 - i 0]} \bigg \} \,.
\label{hadronic dispersion relation}
\end{eqnarray}
Matching the perturbative QCD factorization formula (\ref{QCD result of the point-like photon effect: dispersion form})
of the ``point-like" photon contribution with the corresponding  hadronic representation (\ref{hadronic dispersion relation})
and performing the double Borel transformation with respect to the variables
$(p+q)^2 \to M_1^2$ and $q^2 \to M_2^2$, we derive the sum rules for the coupling constant $g_{M^{\ast} M \gamma}$
of our interest including the light quark mass correction
\begin{eqnarray}
&& f_P \, f_V \, \mu_P \,m_V \,   g_{M^{\ast} M \gamma}^{(\rm per)} \,
{\rm exp} \left [- \left ( {m_V^2 \over M_1^2}  + {m_P^2 \over M_2^2} \right ) \right ] \nonumber \\
&& =  \left ( {N_c \over 4 \, \pi^2} \right ) \, m_Q \, \iint \limits_{\Sigma} d s_1 \, d s_2  \,
{\rm exp} \left [- \left ( {s_1 \over M_1^2}  + {s_2 \over M_2^2} \right ) \right ] \, \rho^{(\rm per)}(s_1, s_2)  \,,
\label{original QCDSR for the point-like contribution}
\end{eqnarray}
where  $\mu_P = m_P^2 / ( m_Q + m_q)$ and the appearance of the integration boundary $\Sigma$
in the $(s_1, s_2)$  plane arises from subtracting the continuum contributions
with the parton-hadron duality ansatz.
Since the  Borel parameters $M_1^2$ and $M_2^2$ are expected to be of similar size numerically,
we will simplify the generic  expression of the LCSR (\ref{original QCDSR for the point-like contribution})
by  making the approximation $M_1^2=M_2^2 \equiv M^2$ in what follows.
Employing the triangle duality region $s_1 + s_2  \leq 2 \, s_0$ as suggested in
\cite{Belyaev:1994zk,Khodjamirian:1999hb} and introducing further the Jacobi transformation
\begin{eqnarray}
s_1 + s_2 = 2 \, s\,, \qquad v={s_2 \over s_1 + s_2} \,,
\end{eqnarray}
we obtain the tree-level LCSR for the  ``point-like" photon contribution
\begin{eqnarray}
f_P \, f_V \, \mu_P  \, m_V \,  g_{M^{\ast} M \gamma}^{(\rm per)}  \,
&=& - \left ( {N_c \over 4 \, \pi^2} \right ) \, m_Q \, \int_{ m_Q^2}^{ s_{0}} d s  \,\,
 {\rm exp} \left [-  {2 \,  s - m_V^2 - m_P^2 \over M^2} \right ]  \nonumber \\
&&   \times \, \bigg \{ e_Q \, \left [ (1-r_q) \, \left (1- {m_Q^2 \over s} \right )
+ \ln {m_Q^2 \over s} \right ]  - e_q \,  \left (1- {m_Q^2 \over s} \right )   \bigg \} \,,
\end{eqnarray}
which can be  reduced further by integrating over the variable $s$ analytically
\begin{eqnarray}
f_P \, f_V \, \mu_P  \, m_V \,  g_{M^{\ast} M \gamma}^{(\rm per)}  \,
&=& \left ( {N_c \over 8 \, \pi^2} \right ) \, m_Q^3 \,  {\rm exp} \left [ {m_V^2 + m_P^2 \over M^2} \right ] \,
\nonumber \\
&& \times  \bigg \{ \left [ {\rm Ei} \left (- {2 \, s_0 \over M^2} \right )
-  {\rm Ei} \left (- {2 \, m_Q^2  \over M^2} \right )  \right ]
\left [ e_Q \, \left (  2 (1- r_q) +  {M^2 \over m_Q^2}  \right )  - 2 \, e_q  \right ]
\nonumber \\
&& \hspace{0.5 cm} + {M^2 \over m_Q^2} \, \left [ {\rm exp} \left (- {2 \, s_0 \over M^2} \right )
-  {\rm exp} \left (- {2 \, m_Q^2  \over M^2} \right )  \right ] \left [ e_Q \left (1- r_q \right ) - e_q  \right ]
\nonumber \\
&& \hspace{0.5 cm} - {M^2 \over m_Q^2} \, {\rm exp} \left (- {2 \, s_0 \over M^2} \right ) \,
\left [e_Q \, \ln {s_0 \over m_Q^2} \right ]  \bigg \}   \,,
\label{sum rules for perturbative contributions}
\end{eqnarray}
with the exponential integral ${\rm Ei}(x)$ defined as
\begin{eqnarray}
{\rm Ei}(x)= - \int_{-x}^{\infty} \, d t \,\, {e^{-t} \over t} \,, \qquad (x \neq 0)\,.
\end{eqnarray}
Several remarks on the obtained sum rules (\ref{sum rules for perturbative contributions})
for the perturbative contribution to the  coupling $g_{M^{\ast} M \gamma}$ with the light quark mass correction are in order.

\begin{itemize}

\item{It would be straightforward  to explore the heavy-quark mass dependence
of all separate terms entering the obtained sum rules (\ref{sum rules for perturbative contributions})
based upon the power counting scheme \cite{Ball:1997rj,Khodjamirian:1998vk}
\begin{eqnarray}
m_V \sim m_P \sim m_Q + \Lambda_{\rm QCD} \,, \qquad
s_0 \sim  (m_Q + \omega_0)^2 \,,  \qquad
f_P \sim f_V \sim {\Lambda_{\rm QCD}^{3/2} / m_Q^{1/2}} \,,
\label{power counting scheme of parameters}
\end{eqnarray}
where the $m_Q$-independent parameter $\omega_0$ is of order $1 \, {\rm GeV}$.
Expanding  (\ref{sum rules for perturbative contributions})
in terms of the electrical charges of the heavy and light quarks
\begin{eqnarray}
g_{M^{\ast} M \gamma}^{(\rm per)} =   e_Q \, g_{M^{\ast} M \gamma}^{(\rm per, \, I)}
+  e_Q \, r_q \,  g_{M^{\ast} M \gamma}^{(\rm per, \, II)}
+ e_q \, g_{M^{\ast} M \gamma}^{(\rm per, \, III)}  \,,
\end{eqnarray}
we can immediately establish the asymptotic scaling laws
\begin{eqnarray}
g_{M^{\ast} M \gamma}^{(\rm per, \, I)} \sim {1 \over m_Q}  \, \left ( {\omega_0 \over \Lambda_{\rm QCD}} \right )^3  \,,
\qquad
g_{M^{\ast} M \gamma}^{(\rm per, \, II)} \sim g_{M^{\ast} M \gamma}^{(\rm per, \, III)}
\sim {1 \over \Lambda_{\rm QCD}}  \, \left ( {\omega_0 \over \Lambda_{\rm QCD}} \right )^2   \,,
\end{eqnarray}
in agreement with the HH$\chi$PT predictions \cite{Amundson:1992yp,Cho:1992nt,Cheng:1992xi}.
}

\item{The substantial cancellation of two different pieces from the photon radiation off the charm and light quarks
can be naturally expected for the magnetic couplings $D_{(s)}^{\ast +} \, D_{(s)}^{+} \, \gamma$.
Consequently, it remains interesting to investigate whether the higher-order QCD corrections to the ``point-like" photon contributions
(see \cite{Beilin:1984pf} for discussions in the context of $J/\psi \to \eta_c \, \gamma$) could alleviate such numerical cancellation.}

\end{itemize}

\section{The twist-two LCSR for the resolved photon  effect }
\label{section: leading-twist hadronic photon effect}

We will proceed to compute the twist-two hadronic photon corrections  to the radiative heavy meson decay
form factors at NLL by employing the LCSR approach.
To this end, we need to establish the corresponding hard-collinear factorization formula for the vacuum-to-photon
correlation function  (\ref{defnition: correlation function}) at ${\cal O}(\alpha_s)$ with the evanescent operator approach
and then implement the RG resummation for enhanced logarithms of the hard-to-collinear scale ratio
with the two-loop evolution equation of the twist-two photon LCDA.

\subsection{Hard-collinear factorization at LO in QCD}

\begin{figure}
\begin{center}
\includegraphics[width=0.45 \columnwidth]{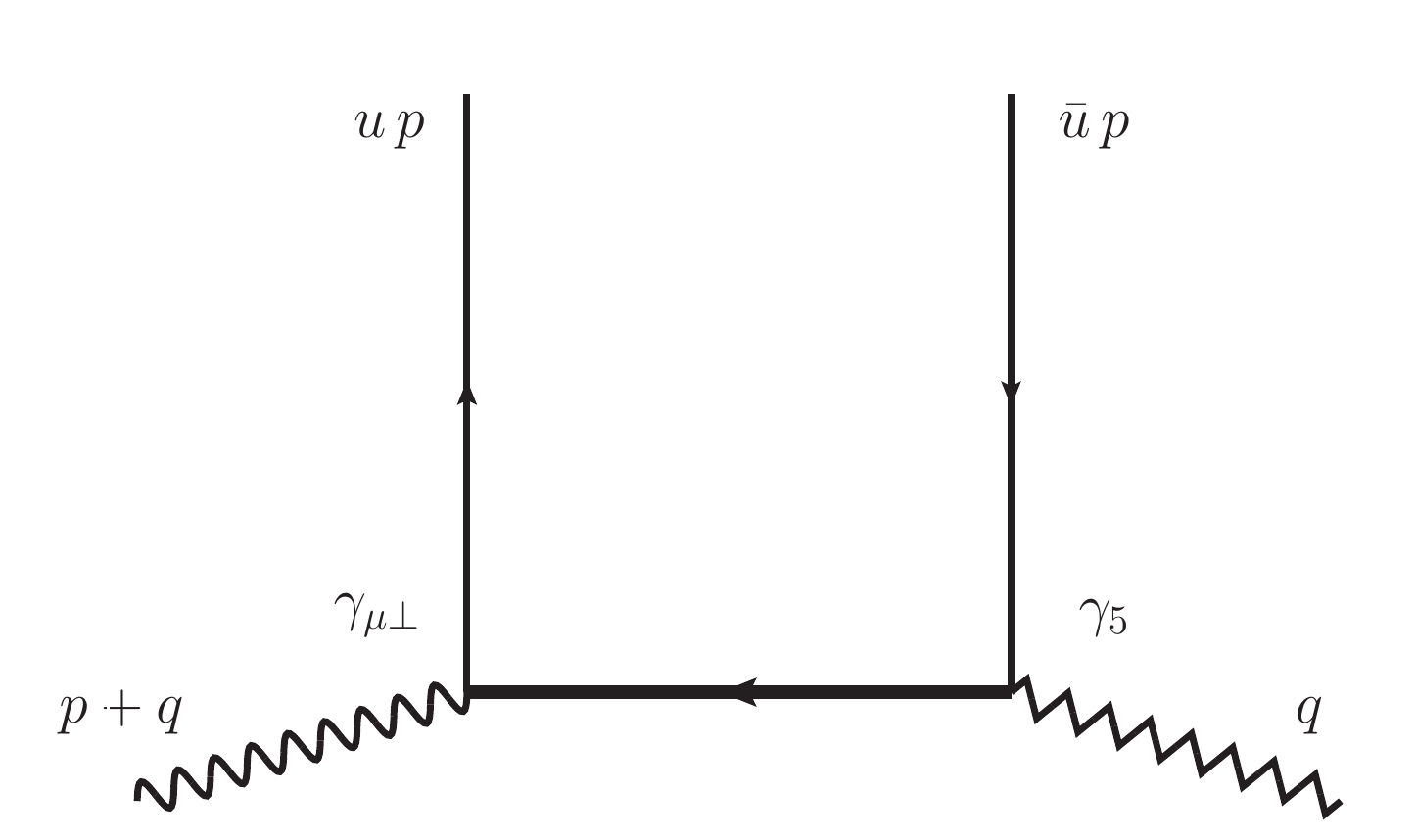}
\vspace*{0.1cm}
\caption{Diagrammatic representation of the  four-point QCD matrix element $F_{\mu}(p, q, u)$
defined by (\ref{defnition: partonic matrix element}) at tree level.}
\label{fig: twist-2 hadronic photon effect at LO}
\end{center}
\end{figure}

Following the standard strategy, the perturbative QCD factorization formula for the correlation function
(\ref{defnition: correlation function}) can be established by inspecting the partonic matrix element
\begin{eqnarray}
F_{\mu}(p, q, u) = \int d^4 x \, e^{-i (p+q) \cdot x} \,
\langle q(u \, p) \, \bar q(\bar u \, p) |  {\rm T }  \left \{  \bar q(x) \gamma_{\mu \perp} Q(x),
\bar Q(0)  \gamma_5 q(0) \right \}  | 0 \rangle  \,,
\label{defnition: partonic matrix element}
\end{eqnarray}
where the dimensionless parameter $u \in [0, 1]$ corresponds to the momentum fraction carried by the collinear quark
and $\bar u \equiv 1-u$.
Evaluating the leading-order Feynman diagram depicted
in figure \ref{fig: twist-2 hadronic photon effect at LO} gives rise to
the tree-level QCD amplitude
\begin{eqnarray}
F_{\mu}^{(0)}(p, q, u) &=& -{i \over 2} \, {\bar n \cdot q \over \bar u (p+q)^2 + u q^2 - m_Q^2 +i 0} \,\,
\bar q(u \, p) \, \gamma_{\mu \perp}  \not \! n \, \gamma_5 \, q(\bar u \, p) \nonumber \\
&=&  -{i \over 2} \, {\bar n \cdot q \over \bar u^{\prime} (p+q)^2 + u^{\prime} q^2 - m_Q^2 +i 0} \,\,
\ast \langle O_{1, \, \mu}(u, u^{\prime}) \rangle^{(0)}  \,,
\end{eqnarray}
where the asterisk represents the convolution integral over the variable $ u^{\prime}$
and the quark matrix element $\langle O_{1, \, \mu}(u, u^{\prime}) \rangle$
of the two-body light-ray operator is given by
\begin{eqnarray}
\langle O_{1, \, \mu}(u, u^{\prime}) \rangle =
\langle q(u \, p) \, \bar q(\bar u \, p) | \hat O_{1,\, \mu}(u^{\prime}) | 0 \rangle
= \bar \xi(u \, p) \, \gamma_{\mu \perp}  \not \! n \, \gamma_5 \, \xi(\bar u \, p)  \,
\delta(u - u^{\prime})+ {\cal O}(\alpha_s) \,.
\end{eqnarray}
The collinear operator  with a given Dirac structure $\Gamma_j$ in momentum space is defined as \cite{Beneke:2005gs}
\begin{eqnarray}
\hat O_{j, \, \mu}(u^{\prime}) =  { n \cdot p \over 2 \pi} \, \int d \tau e^{- i u^{\prime} \, n \cdot p \, \tau }  \,
(\bar \xi \, W_c)(\tau \, n) \, \Gamma_j \, (W_c^{\dagger} \, \xi)(0), \qquad
\Gamma_1 =  \gamma_{\mu \perp}  \not \! n \,\, \gamma_5 \,,
\end{eqnarray}
where the collinear gauge invariance is satisfied by introducing the light-like Wilson line
\begin{eqnarray}
W_c(x) = {\rm P} \, \left \{ {\rm exp}
\left [  i \, g_s \, \int_{-\infty}^{0} d s \,\, n \cdot A_c(x + s \, n) \right] \right  \}   \,.
\end{eqnarray}

Implementing the infrared subtraction for the renormalized matrix element $F_{\mu}(p, q, u)$
can be further achieved  with the aid of the collinear operator basis
\begin{eqnarray}
\hat O_{1, \, \mu} =  \hat O_{2, \, \mu} + \hat O_{E, \, \mu}\,,
\qquad \Gamma_2 = {n^{\nu} \over 2} \, \epsilon_{\mu \nu \rho \tau} \, \sigma^{\rho \tau},
\qquad \Gamma_E = \gamma_{\mu \perp}  \not \! n \,\, \gamma_5
- {n^{\nu} \over 2} \, \epsilon_{\mu \nu \rho \tau} \, \sigma^{\rho \tau}\,,
\end{eqnarray}
where the evanescent operator $\hat O_{E, \, \mu}$ vanishes in the four-dimensional space-time.
Taking advantage of the perturbative  matching relation
\begin{eqnarray}
F_{\mu}(p, q, u) = \sum_{i=2, \, E} H_i((p+q)^2, q^2, u^{\prime}) \ast  \langle O_{i, \, \mu}(u, u^{\prime}) \rangle \,,
\label{matching condition}
\end{eqnarray}
we can readily obtain the short-distance coefficient functions at tree level
\begin{eqnarray}
H_2^{(0)} = H_E^{(0)}  = -{i \over 2} \, {\bar n \cdot q \over \bar u^{\prime} (p+q)^2 + u^{\prime} q^2 - m_Q^2 +i 0} \,.
\end{eqnarray}
It is then straightforward to derive the hard-collinear factorization formula
for the vacuum-to-photon correlation function (\ref{defnition: correlation function})
at leading power in $\Lambda_{\rm QCD}/m_Q$
\begin{eqnarray}
\Pi_{\mu, \, \rm LO}^{\rm tw2}(p, q) = - g_{\rm em} \, e_q \, \chi(\mu) \, \langle \bar q q \rangle(\mu)  \,
\epsilon_{\mu \, p \, q  \,\eta^{\ast}} \,
\int_0^1 d u \frac{\phi_{\gamma}(u, \mu)} {\bar u (p+q)^2 + u q^2 - m_Q^2 +i 0} \,,
\label{twist-2 factorization formula at LO}
\end{eqnarray}
by employing the definition of the twist-two photon distribution amplitude
\begin{eqnarray}
&& \langle  \gamma(p, \eta^{\ast})  | (\bar \xi \, W_c)(\tau \, n) \, \sigma_{\alpha \beta} \, (W_c^{\dagger} \, \xi)(0) | 0 \rangle
\nonumber \\
&& = - i \, g_{\rm em} \, e_q \, \chi(\mu) \, \langle \bar q q \rangle(\mu) \,
(p_{\beta} \eta_{\alpha}^{\ast} -  p_{\alpha} \eta_{\beta}^{\ast} ) \,
\int_0^1 d u \, e^{i \, u \, n \cdot p \, \tau} \, \phi_{\gamma}(u, \mu) \,.
\end{eqnarray}
We proceed to write down the dispersion representation of the factorization formula
(\ref{twist-2 factorization formula at LO})
\begin{eqnarray}
\Pi_{\mu, \, \rm LO}^{\rm tw2}(p, q) &=& - g_{\rm em} \, e_q \, \chi(\mu) \, \langle \bar q q \rangle(\mu)  \,
\epsilon_{\mu \, p \, q  \,\eta^{\ast}} \,\nonumber \\
&&  \times \, \int_{m_Q^2}^{\infty} ds_1 \,  \int_{m_Q^2}^{\infty} ds_2 \,
\frac{\rho^{\rm tw2}_{0}(s_1, s_2)}{[s_1 - (p+q)^2 - i 0][s_2 - q^2 - i 0]} \,,
\end{eqnarray}
where the LO double spectral density \cite{Belyaev:1994zk}
\begin{eqnarray}
\rho^{\rm tw2}_{0}(s_1, s_2) =  \sum_k \, {(-1)^{k+1} \, a_k(\mu)\over  \Gamma(k+1) } \,
(s_1 - m_Q^2)^k  \, \delta^{(k)}(s_1-s_2)\,,
\end{eqnarray}
is obtained by expanding the  twist-two photon LCDA
in terms of the polynomials in $u$
\begin{eqnarray}
\phi_{\gamma}(u, \mu) = \sum_k a_k(\mu) \, u^k \,.
\end{eqnarray}
The resulting LCSR for the resolved photon contribution to the magnetic coupling $g_{M^{\ast} M \gamma}$
is given by
\begin{eqnarray}
f_P \, f_V \, \mu_P  \, m_V \,  g_{M^{\ast} M \gamma}^{(\rm tw2, \, LO)}  \,
&=& -  e_q \, \chi(\mu) \, \langle \bar q q \rangle(\mu) \,
\phi_{\gamma}\left ({1 \over 2}, \mu \right ) \,
\int_{m_Q^2}^{s_0}  ds \,\,  {\rm exp} \left [  {m_V^2 + m_P^2 - 2 \, s  \over M^2}\right ].
\label{twist-2 LCSR at LO}
\end{eqnarray}
Applying the default power counting scheme displayed in (\ref{power counting scheme of parameters})
we find
\begin{eqnarray}
g_{M^{\ast} M \gamma}^{(\rm tw2, \, LO)} \sim {1 \over \Lambda_{\rm QCD}} \, \left ( {\omega_0 \over \Lambda_{\rm QCD}} \right ),
\label{scaling law of the twist-2 effect}
\end{eqnarray}
which is suppressed by a factor of ${\Lambda_{\rm QCD} / \omega_0}$ compared with the ``point-like" photon effect.

\subsection{Hard-collinear factorization at NLO in QCD}

\begin{figure}
\begin{center}
\includegraphics[width=1.0 \columnwidth]{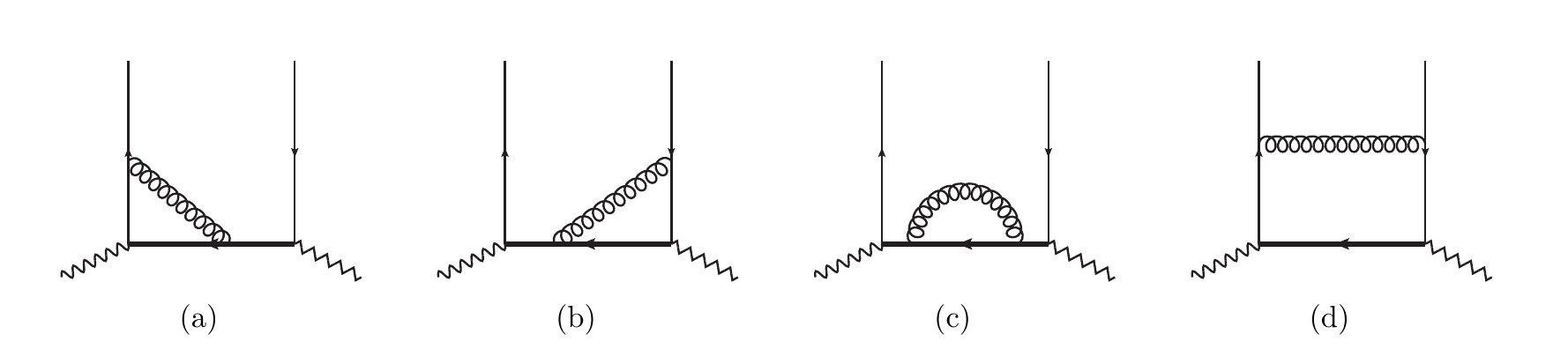}
\vspace*{0.1cm}
\caption{Diagrammatic representation of the  four-point QCD matrix element $F_{\mu}(p, q, u)$
defined by (\ref{defnition: partonic matrix element}) at one loop.}
\label{fig: twist-2 hadronic photon effect at NLO}
\end{center}
\end{figure}

We are now in a position to demonstrate the hard-collinear factorization formula
for the vacuum-to-photon correlation function (\ref{defnition: correlation function})
at NLO in QCD. The determination of the corresponding perturbative matching coefficient
can be achieved by evaluating the one-loop diagrams displayed in
figure \ref{fig: twist-2 hadronic photon effect at NLO} with the aid of (\ref{matching condition}).
Evaluating the QCD correction to the vector-meson vertex diagram shown in
figure \ref{fig: twist-2 hadronic photon effect at NLO} (a) leads to
\begin{eqnarray}
F_{\mu}^{1, \, (a)} &=& {g_s^2 \, C_F \over \bar u (p+q)^2 + u q^2 - m_Q^2} \,
\int {d^D \ell \over (2 \, \pi)^D} \,
{1 \over [(\ell + u \, p)^2  + i0] [(\ell - \bar u \, p - q)^2 -m_Q^2 + i0] [\ell^2 + i 0]}  \nonumber \\
&& \bar q(u \, p) \, \gamma_{\nu}  \, (u \not \! p + \not \! \ell) \, \gamma_{\mu \perp} \,
(\not \! \ell  - \bar u \not \! p \, - \not \! q + m_Q) \, \gamma^{\nu} \,
( -\bar u \not \! p \, - \not \! q + m_Q) \, \gamma_5 \,\, q(\bar u \, p)\,.
\end{eqnarray}
Applying the power counting scheme for the external momenta presented in (\ref{power counting scheme}),
we can identify the leading power contributions of the scalar Feynman integral
\begin{eqnarray}
I_a = \int {d^D \ell \over (2 \, \pi)^D} \,
{1 \over [(\ell + u \, p)^2  + i0] [(\ell - \bar u \, p - q)^2 -m_Q^2 + i0] [\ell^2 + i 0]} \,
\end{eqnarray}
from the hard and collinear regions.
Employing the strategy of regions constructed in \cite{Beneke:1997zp,Smirnov:2002pj},
the collinear region integral at leading power in the heavy quark expansion can be reduced as
\begin{eqnarray}
I_a^{(c)} = - \int {d^D \ell \over (2 \, \pi)^D} \,
{1 \over [n \cdot (\ell + u \, p) \,\, \bar n \cdot \ell + \ell_{\perp}^2  + i0]
[n \cdot (\ell - \bar u \, p - q) \, \bar n \cdot q  + m_Q^2 - i0] [\ell^2 + i 0]} \,,
\hspace{0.5 cm}
\end{eqnarray}
which results in a vanishing contribution with the dimensional regularization scheme
and will be also cancelled out by the corresponding collinear subtraction term with an arbitrary
regularization scheme.
Taking advantage of the expressions for the  hard-region integrals collected in
Appendix \ref{appendix: Feynman integrals at one loop} as well as the naive dimensional regularization (NDR)
scheme for the Dirac matrix $\gamma_5$,  the one-loop correction to the vector-meson vertex  diagram can be computed as
\begin{eqnarray}
F_{\mu}^{1, \, (a)} &=&  {\alpha_s \, C_F \over 4 \, \pi }  \,
\bigg \{  \bigg  [ {1 \over \epsilon} + \ln {\mu^2 \over m_Q^2} -\ln \left [(1-r_1) (1-r_2) \right ]
- {r_1  - 2\, r_2 +1 \over 2 \, r_1 \, (1-r_2)} + {3 \over 2} \bigg ]  \nonumber \\
&&  \times \, \left [ { 2 \, (1-r_2) \over r_2 - r_1 } \, \ln {1 - r_1 \over 1 - r_2}  - 1 \right ]
+ {2 \, (1-r_2) \over r_1 -r_2}  \, \left [ {\rm Li_2}(r_1) - {\rm Li_2}(r_2) \right ]
\nonumber \\
&&  - \left ( {1 - 2 \, r_2 \over r_1 \, r_2 }  + 1 \right ) \, \ln (1-r_2)
- {r_1 -2 \, r_2 + 1 \over 2 \, r_1 \, (1-r_2)}
 - {5 \over 2} \bigg \} \, F_{\mu}^{0} \,,
\end{eqnarray}
where we have introduced the following dimensionless variables
\begin{eqnarray}
r_1 \equiv  { (\bar u \, p+ q)^2 / m_Q^2}  \,, \qquad  r_2 \equiv  { (p + q)^2 / m_Q^2} \,.
\end{eqnarray}

Along the same vein, we can write down the one-loop amplitude for the
QCD correction to the pseudoscalar-meson vertex diagram displayed in
figure \ref{fig: twist-2 hadronic photon effect at NLO} (b)
\begin{eqnarray}
F_{\mu}^{1, \, (b)} &=& {g_s^2 \, C_F \over \bar u (p+q)^2 + u q^2 - m_Q^2} \,
\int {d^D \ell \over (2 \, \pi)^D} \,
{1 \over [(\ell - \bar u \, p)^2  + i0] [(\ell - \bar u \, p - q)^2 -m_Q^2 + i0] [\ell^2 + i 0]}  \nonumber \\
&& \bar q(u \, p) \, \gamma_{\mu \perp}  \, (- \bar u \not \! p \, -  \not \! q  + m_Q) \, \gamma_{\nu} \,
(\not \! \ell  - \bar u \not \! p \, - \not \! q + m_Q) \, \gamma_5 \,\,
( \not \! \ell  -  u \not \! p) \,  \gamma^{\nu}  \, q(\bar u \, p)\,.
\end{eqnarray}
Apparently, only the hard and collinear loop-momentum regions can generate
the leading power contributions to $F_{\mu}^{1, \, (b)}$ with the power
counting scheme (\ref{power counting scheme}).
Implementing the Dirac algebra reduction with the equation of motion for the quark field
and  performing the loop-momentum integration with the formulae displayed in
Appendix \ref{appendix: Feynman integrals at one loop}, we obtain
\begin{eqnarray}
F_{\mu}^{1, \, (b)} &=&  {\alpha_s \, C_F \over 2 \, \pi }  \,
\bigg \{  \bigg  [ {1 \over \epsilon} + \ln {\mu^2 \over m_Q^2} -\ln \left [(1-r_1) (1-r_3) \right ]
- {r_3  - 2 \, r_1 +1 \over  r_1 \, (1 - r_3)} \bigg ]  \nonumber \\
&&  \times \, \left [ {  1-r_3 \over r_3 - r_1 } \, \ln {1 - r_1 \over 1 - r_3}  + 1 \right ]
+ {1-r_3 \over r_1 -r_3}  \, \left [ {\rm Li_2}(r_1) - {\rm Li_2}(r_3) \right ]
\nonumber \\
&&  +  \left ( {r_3 + 1  \over r_1 \, r_3 }  + 1 \right ) \, \ln (1-r_3)
- {r_3 -2 \, r_1 + 1 \over  \, r_1 \, (1-r_3)} + 1 \bigg \} \, F_{\mu}^{0} \,,
\end{eqnarray}
where we have defined the convention $r_3 \equiv q^2/m_Q^2$ for brevity.

The self-energy correction to the heavy-quark propagator shown in
figure \ref{fig: twist-2 hadronic photon effect at NLO} (c) is
evidently free of the collinear divergence and
can be readily computed as follows
\begin{eqnarray}
F_{\mu}^{1, \, (c)} &=& {g_s^2 \, C_F \over \bar u (p+q)^2 + u q^2 - m_Q^2} \,
\int {d^D \ell \over (2 \, \pi)^D} \,
{1 \over [(\ell - \bar u \, p - q)^2 -m_Q^2 + i0] [\ell^2 + i 0]}  \nonumber \\
&& \bar q(u \, p) \, \gamma_{\mu \perp}  \, (- \bar u \not \! p \,  -  \not \! q  + m_Q) \, \gamma_{\nu} \,
(\not \! \ell  - \bar u \not \! p \, - \not \! q + m_Q) \,  \gamma^{\nu}  \, \,
(- \bar u \not \! p \, -  \not \! q + m_Q) \, \gamma_5 \, q(\bar u \, p)\,, \nonumber \\
&=& - {\alpha_s \, C_F \over 4 \, \pi} \, \bigg \{ {7 - r_1 \over 1- r_1} \,
\left [ {1 \over \epsilon} +\ln {\mu^2 \over m_Q^2} - \ln (1-r_1) + 1 \right ]
\nonumber \\
&& \hspace{1.8 cm} - \frac{1}{1-r_1} \bigg[ \frac{1 - 7 \, r_1}{r_1^2}\ln(1-r_1) + \frac{1}{r_1} - 3 \bigg] \bigg\}
\, F_{\mu}^{0} \,,
\end{eqnarray}
by employing the results of the one-loop Feynman integrals collected  in Appendix \ref{appendix: Feynman integrals at one loop}.

The one-loop QCD amplitude from the box diagram displayed in figure
\ref{fig: twist-2 hadronic photon effect at NLO} (d) is given by
\begin{eqnarray}
F_{\mu}^{1, \, (d)} &=& g_s^2 \, C_F \, \int {d^D \ell \over (2 \, \pi)^D} \,
{1 \over [(\ell - \bar u \, p)^2  + i0] [(\ell + u \, p)^2  + i0]
[(\ell - \bar u \, p - q)^2 -m_Q^2 + i0] [\ell^2 + i 0]}  \nonumber \\
 \nonumber \\
&& \bar q(u \, p) \,  \gamma_{\nu} \, (\not \! \ell  \, +  u  \not \! p) \, \gamma_{\mu \perp}  \,
(\not \! \ell  - \bar u \not \! p \, - \not \! q + m_Q) \, \gamma_5 \,
(\not \! \ell - \bar u \not \! p) \,  \gamma^{\nu}  \,  q(\bar u \, p)\,,
\end{eqnarray}
which can be further reduced with the NDR scheme of $\gamma_5$
\begin{eqnarray}
F_{\mu}^{1, \, (d)} &=& i \, g_s^2 \, C_F \, \int {d^D \ell \over (2 \, \pi)^D} \,
{m_Q^2 \, (1-r_1) \over [(\ell - \bar u \, p)^2  + i0] [(\ell + u \, p)^2  + i0]
[(\ell - \bar u \, p - q)^2 -m_Q^2 + i0] [\ell^2 + i 0]}  \nonumber \\
 \nonumber \\
&& (D-4) \, \left \{ {D-4 \over D-2} \, \ell_{\perp}^2
+  {\bar n \cdot \ell \over \bar n \cdot q} \,
\left[ \ell^2 - \bar n \cdot \ell \,\, n \cdot (\bar u \, p + q)  \right ] \right \}
\, F_{\mu}^{0} \,.
\end{eqnarray}
Evaluating the tensor four-point integrals with the dimensional regularization scheme
implies that  the obtained amplitude $F_{\mu}^{1, \, (d)}$ only contributes at ${\cal O}(\epsilon)$
and therefore vanishes at $D=4$, in agreement with the analogous observation for the hadronic
photon corrections to the radiative leptonic $B$-meson decays \cite{Wang:2018wfj}.

Adding up different pieces together we can derive the one-loop contribution to the  QCD
matrix element $F_{\mu}$ as follows
\begin{eqnarray}
F_{\mu}(p, q, u) = \sum_{i=2, \, E} T_i^{(1)}((p+q)^2, q^2, u^{\prime}) \ast  \langle O_{i, \, \mu}(u, u^{\prime}) \rangle^{(0)} \,,
\label{one-loop QCD amplitude at twist-2}
\end{eqnarray}
where the explicit expressions of the hard amplitudes $T_i^{(1)}$ with the NDR scheme of $\gamma_5$ read
\begin{eqnarray}
&& T_i^{(1)} ((p+q)^2, q^2, u^{\prime})  \nonumber\\
&& =  { \alpha_s \, C_F \over 4 \, \pi} \, \bigg \{ (-2) \,
\Big [ {1 -r_2 \over r_1-r_2} \, \ln {1-r_1 \over 1-r_2}
+ {1 -r_3 \over r_1-r_3}  \, \ln {1-r_1 \over 1-r_3} + {3 \over 1-r_1}\Big ]\,
\Big ({1 \over \epsilon} + \ln {\mu^2 \over m_Q^2} \Big ) \nonumber \\
&&+2\,\Big[\Big(\frac{1-r_2}{r_1-r_2}+\frac{1-r_3}{r_1-r_3}\Big)\,{\rm Li}_2(r_1) -\frac{1-r_2}{r_1-r_2}\,{\rm Li}_2(r_2) -\frac{1-r_3}{r_1-r_3}\,{\rm Li}_2(r_3) \Big]\nonumber\\
&&+2\,\Big[\Big(\frac{1-r_2}{r_1-r_2}+\frac{1-r_3}{r_1-r_3}\Big)\,\ln^2(1-r_1)-\frac{1-r_2}{r_1-r_2}\,\ln^2(1-r_2) -\frac{1-r_3}{r_1-r_3}\,\ln^2(1-r_3) \Big]\nonumber\\
&&+\Big[\frac{1-r_1}{r_1}\Big(\frac{1-r_1-2\,r_2}{r_1-r_2}-\frac{2\,(1-r_1+r_3)}{r_1-r_3}\Big)\,
+\frac{1-6\,r_1}{r_1^2}+1\Big]\,\ln(1-r_1) \nonumber\\
&&+\frac{1-9\,r_1}{r_1\,(1-r_1)} -\frac{(1-r_2)\,(1-3\,r_2)}{r_2\,(r_1-r_2)}\,\ln(1-r_2)
+\frac{2\,(1-r_3)}{r_3\,(r_1-r_3)}\,\ln(1-r_3) -3 \bigg \} \, H_i^{(0)} \,,
\label{one-loop QCD amplitude at twist-2}
\end{eqnarray}
with the variable $u$ in the definition of $r_1$ replaced by $u^{\prime}$.
Expanding all the quantities in the operator matching relation (\ref{matching condition})
at the one-loop accuracy yields
\begin{eqnarray}
\sum_{i} T_i^{(1)}((p+q)^2, q^2, u^{\prime}) \ast  \langle O_{i, \, \mu}(u, u^{\prime}) \rangle^{(0)} &=
&\sum_{i} \big [  H_i^{(1)}((p+q)^2, q^2, u^{\prime}, \mu) \ast  \langle O_{i, \, \mu}(u, u^{\prime}) \rangle^{(0)}
\nonumber \\
&& \hspace{0.2 cm} + \,  H_i^{(0)}((p+q)^2, q^2, u^{\prime}) \ast  \langle O_{i, \, \mu}(u, u^{\prime}, \mu) \rangle^{(1)} \big ].
\hspace{0.8 cm}
\end{eqnarray}
The ultraviolet renormalized collinear matrix elements $\langle O_{i, \, \mu} \rangle^{(1)}$ are given by
\begin{eqnarray}
\langle O_{i, \, \mu} \rangle^{(1)} =  \left [ M_{i j}^{(1), \, R} + Z_{i j}^{(1)} \right ] \,
\langle O_{i, \, \mu} \rangle^{(0)}  \,,
\end{eqnarray}
where $ M_{i j}^{(1), \, R}$ and $ Z_{i j}^{(1)}$ represent the
bare matrix elements computed from the infrared regularization scheme $R$
and the renormalization factors for subtracting the ultraviolet divergences
at one loop, respectively.
Following the prescriptions constructed in \cite{Beneke:2005vv},
the one-loop hard matching coefficient $H_2^{(1)}$ of the physical operator  $\hat{O}_{2, \, \mu}$
can be determined as
\begin{eqnarray}
H_2^{(1)} = T_2^{(1)} - H_2^{(0)} \ast Z_{22}^{(1)} + H_E^{(0)} \ast   M_{E 2}^{(1), \, \rm off} \,,
\end{eqnarray}
where the infrared subtraction term $H_E^{(0)} \ast   M_{E 2}^{(1), \, \rm off}$
arises from  the  mixing of  $\hat{O}_{E, \, \mu}$ into  $\hat{O}_{2, \, \mu}$.
Computing the one-loop matrix element of the evanescent operator explicitly leads to \cite{Wang:2018wfj}
\begin{eqnarray}
H_E^{(0)} \ast   M_{E 2}^{(1), \, \rm off}  = 0 \,,
\end{eqnarray}
from which we can immediately write down the desired short-distance matching coefficient
\begin{eqnarray}
H_2^{(1)} =  T_2^{(1), \, \rm reg} \,.
\end{eqnarray}
It is apparent that $T_2^{(1), \, \rm reg}$ corresponds to the regularized terms of
the NLO QCD matrix element $T_2^{(1)}$ displayed in (\ref{one-loop QCD amplitude at twist-2}).

The one-loop hard-collinear factorization formula for the vacuum-to-$B$-meson correlation
function at the twist-two accuracy is then given by
\begin{eqnarray}
\Pi_{\mu, \, \rm NLO}^{\rm tw2}(p, q) &=& - g_{\rm em} \, e_q \, \chi(\mu) \, \langle \bar q q \rangle(\mu)  \,
\epsilon_{\mu \, p \, q  \,\eta^{\ast}} \,
\int_0^1 d u \frac{\phi_{\gamma}(u, \mu)} {\bar u (p+q)^2 + u q^2 - m_Q^2 +i 0} \,  \nonumber \\
&& \left[ 1 +  \frac{H_2^{(1)}((p+q)^2, q^2, u, \mu)}{H_2^{(0)}((p+q)^2, q^2, u)} \right ]
+ {\cal O}(\alpha_s^2) \,.
\label{one-loop factorization formula at twist 2}
\end{eqnarray}
In order to demonstrate the factorization-scale independence of the obtained expression
for the correlation function (\ref{defnition: correlation function}),
we need to distinguish the QCD renormalization scale $\nu$ for the pseudoscalar interpolating current
with the hard-collinear factorization scale $\mu$.
Such distinction can be explicitly achieved by employing the following decomposition
\begin{eqnarray}
H_2^{(1)}((p+q)^2, q^2, u, \mu, \nu) = H_2^{(1)}((p+q)^2, q^2, u, \mu)
+ \delta H_2^{(1)}((p+q)^2, q^2, u, \mu, \nu)  \,,
\end{eqnarray}
where the second term on the right-hand side $ \delta H_2^{(1)}$ satisfies the
RG evolution equation
\begin{eqnarray}
{d \over d \ln \mu} \delta H_2^{(1)}((p+q)^2, q^2, u, \mu, \nu)
=  {\alpha_s(\mu) \over 4 \, \pi} \, \gamma_m^{(0)} \, H_2^{(0)}((p+q)^2, q^2, u) \,,
\qquad  \gamma_m^{(0)} = 6 \, C_F \,,
\end{eqnarray}
and the consistency condition valid to all orders of $\alpha_s$
\begin{eqnarray}
\delta H_2^{(1)}((p+q)^2, q^2, u, \mu, \nu=\mu) = 0\,.
\end{eqnarray}
It is then straightforward to find
\begin{eqnarray}
\delta H_2^{(1)}((p+q)^2, q^2, u, \mu, \nu) =
\left [ {\alpha_s(\mu) \over 4 \, \pi} \, \gamma_m^{(0)} \,  \ln \left ( {\nu \over \mu} \right ) \right ]
\, H_2^{(0)}((p+q)^2, q^2, u)  \,.
\end{eqnarray}
According to the RG evolution equation of the twist-two photon LCDA \cite{Lepage:1979zb}
\begin{eqnarray}
{d  \over d \ln \mu}  \, \left [ \chi(\mu) \, \langle \bar q q \rangle(\mu) \, \phi_{\gamma}(u, \mu)  \right ]
= \int_0^1 d u^{\prime} \, V(u, u^{\prime}) \,
\left [ \chi(\mu) \, \langle \bar q q \rangle(\mu) \, \phi_{\gamma}(u^{\prime}, \mu)  \right ] \,,
\end{eqnarray}
and expanding the renormalization kernel in terms of the strong coupling constant
\begin{eqnarray}
V(u, u^{\prime}) &=&  \sum_{n} \, \left ( {\alpha_s(\mu) \over 4 \pi} \right )^{n+1} \, V^{(n)}(u, u^{\prime} ) \,,
\nonumber \\
 V^{(n)}(u, u^{\prime} ) &=& 2 \, C_F \left [ {1- u \over 1- u^{\prime}} \, {\theta(u-u^{\prime}) \over u - u^{\prime} }
+ {u \over u^{\prime}}\,  {\theta(u^{\prime}-u) \over u^{\prime} - u}   \right ]_{+} - C_F \, \delta(u - u^{\prime}) \,,
\end{eqnarray}
we can readily verify that the perturbative QCD factorization formula (\ref{one-loop factorization formula at twist 2})
for the twist-two hadronic photon contribution to the correlation function  (\ref{defnition: correlation function})
is indeed independent of the factorization scale $\mu$ at one loop as expected. More explicitly, we obtain
\begin{eqnarray}
{d \, \Pi_{\mu, \, \rm NLO}^{\rm tw2} \over d \, \ln \mu} = {\cal O}(\alpha_s^2) \,.
\end{eqnarray}

Apparently, the parametrically large logarithms of $m_Q/\Lambda_{\rm QCD}$ entering
the hard-collinear factorization formula (\ref{one-loop factorization formula at twist 2})
cannot be eliminated by making use of the common value of the factorization scale $\mu$.
The NLL resummation of such logarithms to all orders can be achieved by solving
the evolution equation for the twist-two photon LCDA  at two loops.
Expanding $\phi_{\gamma}(u, \mu)$ in a series of Gegenbauer polynomials \cite{Ball:2002ps}
\begin{eqnarray}
\phi_{\gamma}(u, \mu) = 6 \, u \, (1-u) \, \sum_{n}^{\infty} \, b_{n}(\mu)  \,
C_{n}^{3/2}(2u-1) \,,
\end{eqnarray}
the general solution to the above-mentioned RG equation can be constructed as follows \cite{Mueller:1993hg}
\begin{eqnarray}
\chi(\mu) \, \langle \bar q q \rangle (\mu) \, b_n(\mu) &=& E_{T, n}^{\rm NLO}(\mu, \mu_0) \,
\left [ \chi(\mu_0) \, \langle \bar q q \rangle (\mu_0) \, b_n(\mu_0) \right  ]  \nonumber \\
&& + \, {\alpha_s(\mu) \over 4 \pi} \, \sum_{k=0}^{n-2} \,  E_{T, n}^{\rm LO}(\mu, \mu_0) \,
d_{T, n}^{k}(\mu, \mu_0) \, \left [ \chi(\mu_0) \, \langle \bar q q \rangle (\mu_0) \, b_k(\mu_0) \right ]  \,,
\label{two-loop  evolution of twist-2 photon DA}
\end{eqnarray}
where the manifest expressions of the evolution functions $E_{T, n}^{\rm (N)LO}$ and $d_{T, n}^{k}$
can be found in \cite{Wang:2017ijn}, and $n, \, k$ are non-negative even integers.
The obtained hard-collinear factorization formula for the correlation function (\ref{defnition: correlation function})
with the RG improvement can be further written as
\begin{eqnarray}
\Pi_{\mu, \, \rm NLL}^{\rm tw2}(p, q) &=& - g_{\rm em} \, e_q \, \chi(\mu) \, \langle \bar q q \rangle(\mu)  \,
\epsilon_{\mu \, p \, q  \,\eta^{\ast}} \, \nonumber \\
&& \sum_{n} \, b_n(\mu) \, \left [ G_{n}^{(0)}((p+q)^2, q^2)
+ {\alpha_s(\mu) \, C_F \over 4 \, \pi} \, G_{n}^{(1)}((p+q)^2, q^2, \mu)   \right ],
\label{NLL factorization formula at twist 2}
\end{eqnarray}
where the perturbative kernels $ G_{n}^{(i)} \, (i=0, \, 1)$ are defined by the convolution integrals
\begin{eqnarray}
G_{n}^{(0)}((p+q)^2, q^2) &=& \int_0^1 d u \, \frac{6 \, u \, (1-u) \, C_{n}^{3/2}(2u-1)} {\bar u (p+q)^2 + u q^2 - m_Q^2 +i 0} \,
 \,,  \nonumber \\
G_{n}^{(1)}((p+q)^2, q^2) &=& \int_0^1 d u \, \frac{6 \, u \, (1-u) \, C_{n}^{3/2}(2u-1)} {\bar u (p+q)^2 + u q^2 - m_Q^2 +i 0} \,
\, \frac{H_2^{(1)}((p+q)^2, q^2, u, \mu)}{H_2^{(0)}((p+q)^2, q^2, u)} \,.
\end{eqnarray}

In analogy to the construction of the tree-level LCSR (\ref{twist-2 LCSR at LO})
for the twist-two resolved photon contribution,
we proceed to derive the  spectral representation of the resummation improved
factorization formula (\ref{NLL factorization formula at twist 2})
\begin{eqnarray}
\Pi_{\mu, \, \rm NLL}^{\rm tw2}(p, q) &=& - g_{\rm em} \, e_q \, \chi(\mu) \, \langle \bar q q \rangle(\mu)  \,
\epsilon_{\mu \, p \, q  \,\eta^{\ast}} \, \int_{0}^{\infty} ds_1 \,  \int_{0}^{\infty} ds_2 \,
\frac{1}{[s_1 - (p+q)^2 - i 0][s_2 - q^2 - i 0]} \nonumber \\
&&  \times \, \left [  \rho^{\rm tw2}_{0}(s_1, s_2)
+ {\alpha_s(\mu) \, C_F \over 4 \, \pi} \, \rho^{\rm tw2}_{1}(s_1, s_2)  \right ] \,.
\end{eqnarray}
Following the discussion presented in \cite{Khodjamirian:1999hb},
it turns out to be sufficient to extract the NLO spectral density $\rho^{\rm tw2}_{1}(s_1, s_2)$
with the asymptotic photon distribution amplitude $\phi_{\gamma}^{\rm asy}(u, \mu)=6 \, u \, (1-u)$
by discarding the subdominant effects due to the non-vanishing higher Gegenbauer moments.
Taking advantage of the spectral representations of various complex functions
displayed in Appendix \ref{appendix: master formulae for double spectral densities},
it is then straightforward to derive the analytical expression of the double spectral density
\begin{eqnarray}
m_Q^2 \, \rho^{\rm tw2}_{1}(s_1, s_2) &=&
\big \{ [ \left [\varrho_{\rm I}(r, \sigma)  + \varrho_{\rm II}(r, \sigma) \, \ln r
+ \varrho_{\rm III}(r, \sigma) \, (\ln^2 r - \pi^2) \right ] \,
\delta^{(2)}(r-1) \nonumber \\
&& +  \left [ \varrho_{\rm II}(r, \sigma) + 2 \, \varrho_{\rm III}(r, \sigma)  \, \ln r  \right ] \,
{d^3 \over d r^3} \, \ln|1-r|  \big \}  \, \theta(s_1 - m_Q^2)  \, \theta(s_2 - m_Q^2) \,,
\hspace{0.5 cm}
\end{eqnarray}
where we have introduced the invariant functions
\begin{eqnarray}
\varrho_{\rm I}(r, \sigma) &=& \frac{(r+1)}{\sigma^3}
\Bigg\{6 \,r\,\sigma^2\, \left [ 2 \, \ln \left ( \frac{\mu^2}{m_Q^2} \right )
+ {3 \over 2} \, \ln \left ( \frac{\nu^2}{\mu^2} \right )
+ 3\,{\rm Li}_2 \left (- \frac{\sigma }{r+1} \right )
+ {\rm Li}_2 \left (- \frac{r\, \sigma }{r+1} \right )  \right ]  \nonumber \\
&&  + \, 6 \, r\,\sigma^2\,  \ln \left ( \frac{\sigma}{r+1} \right ) \,
\left [ \ln \left ( \frac{\sigma+r+1 }{r+1} \right ) + \ln \left ( \frac{r \sigma + r + 1 }{r+1} \right ) \right ]  \nonumber\\
&& + \, 3\,(r+1)\,\big[2\, r \,(\sigma+1)+5 \,\sigma+2\big]\, \ln \left ( \frac{\sigma+r+1 }{r+1} \right ) \nonumber\\
&& - \, \frac{3\,r\,\sigma^2}{(r+\sigma+1)(r\sigma+r+1)}\left [9\,r\,\sigma^2+(7\,r+10)\,(r+1)\,\sigma+8\,(r+1)^2 \right ] \,
\ln \left ( \frac{\sigma }{r+1} \right )  \nonumber \\
&& + \, \frac{\sigma}{r+\sigma+1}\,\left [ \big((3+2\,\pi^2)\, r+6\big)\, \sigma^2
+\big((9+2\,\pi^2)\, r-6\big)\, (r+1)\, \sigma-6\, (r+1)^2 \right ] \Bigg\} \nonumber \\
&& + \, 6 \, m_Q^4\, \left [ 4+3\,\ln \left ( \frac{\mu^2}{m_Q^2} \right ) \right ]\,
\frac{d}{dm_Q^2}\,\left [ \frac{1}{m_Q^2}\,\frac{r\,(r+1)}{\sigma} \right ] \,,  \\
\varrho_{\rm II}(r, \sigma) &=&  3 \,r(r+1)\,\left[ \frac{r}{r\,\sigma+r+1}
+ \frac{2}{r+\sigma+1} - \frac{2}{\sigma}\, \ln \left( \frac{\sigma+r+1}{r\,\sigma+r+1} \right ) \right ] \,,
\\
\varrho_{\rm III}(r, \sigma) &=& - \frac{6 \,r\,(r+1)}{\sigma} \,,
\end{eqnarray}
with  two new dimensionless variables defined by
\begin{eqnarray}
r= {s_1 - m_Q^2 \over s_2 - m_Q^2 } \,, \qquad
\sigma = {s_1 \over m_Q^2} + {s_2 \over m_Q^2} - 2 \,.
\label{Jacobi transformation r and sigma}
\end{eqnarray}
The obtained NLL LCSR for the twist-two hadronic photon correction to the radiative
heavy meson decay form factor can be written as
\begin{eqnarray}
&& f_P \, f_V \, \mu_P \,m_V \,   g_{M^{\ast} M \gamma}^{(\rm tw2, \, NLL)} \,
{\rm exp} \left [- \left ( {m_V^2 + m_P^2 \over M^2}  \right ) \right ] \nonumber \\
&& = e_q \, \chi(\mu) \, \langle \bar q q \rangle(\mu)  \,  \iint \limits_{\Sigma} d s_1 \, d s_2  \,
{\rm exp} \left [- \left ( {s_1 + s_2 \over M^2} \right ) \right ]  \,
\left [  \rho^{\rm tw2}_{0}(s_1, s_2)
+ {\alpha_s(\mu) \, C_F \over 4 \, \pi} \, \rho^{\rm tw2}_{1}(s_1, s_2)  \right ]\,.
\hspace{1.0 cm}
\label{NLL LCSR at twist 2 in s1 and s2}
\end{eqnarray}
Since the NLL spectral density $\rho^{\rm tw2}_{1}(s_1, s_2)$ is not concentrated at $s_1=s_2$,
the result of the dispersion integral entering (\ref{NLL LCSR at twist 2 in s1 and s2}) now
depends on the precise shape of the duality region $\Sigma$ in the $(s_1, s_2)$ plane,
in contrast to the tree-level LCSR displayed in  (\ref{twist-2 LCSR at LO}).
Employing the duality boundary $s_1 + s_2 \leq 2 \, s_0$ implemented in \cite{Khodjamirian:1999hb}
and performing the Jacobi transformation introduced in (\ref{Jacobi transformation r and sigma}),
we can readily derive
\begin{eqnarray}
&& f_P \, f_V \, \mu_P(\nu) \,m_V \,   g_{M^{\ast} M \gamma}^{(\rm tw2, \, NLL)} \,
{\rm exp} \left [- \left ( {m_V^2 + m_P^2 \over M^2}  \right ) \right ] \nonumber \\
&& = e_q \, \chi(\mu) \, \langle \bar q q \rangle(\mu)  \,
\bigg \{  - {M^2 \over 2} \, \left [ {\rm exp} \left ( - {2 \, m_Q^2 \over M^2}  \right )
- {\rm exp} \left ( - {2 \, s_0  \over M^2}  \right )  \right ] \,
\phi_{\gamma}\left ({1 \over 2}, \mu \right )   \nonumber \\
&& \hspace{0.5 cm} + \, {\alpha_s(\mu) \, C_F \over 4 \, \pi} \,
\bigg [ \int_{2 \, m_Q^2}^{2 \, s_0} \, ds \, {\rm exp} \left ( - {s  \over M^2}  \right ) \,
{\cal F}^{(\rm tw2)} \left ({s \over m_b^2} -2 \right )   \nonumber \\
&& \hspace{3.0 cm} + \, 3 \,  m_Q^2\, \left (4+3\,\ln\frac{\mu^2}{m_Q^2} \right )  \,
{\rm exp} \left ( - {2\, m_Q^2  \over M^2}  \right )  \bigg ] \bigg \} \,,
\label{final result of the NLL LCSR at twist 2}
\end{eqnarray}
where the explicit expression of ${\cal F}^{(\rm tw2)}(\sigma)$ is given by
\begin{eqnarray}
{\cal F}^{(\rm tw2)}(\sigma) &=& -3 \, \bigg  \{ \ln \left ( {\mu^2 \over m_Q^2} \right )
+ {3 \over 4} \,   \ln \left ( {\nu^2 \over \mu^2} \right ) - {\rm Li}_2 \left ( - \sigma \right )
- {\rm Li}_2 \left ( - \sigma - 1 \right )  + 2 \,  {\rm Li}_2 \left ( - {\sigma \over 2} \right )
+ {\pi^2 \over 12} \nonumber \\
&& + \, \ln \left ( \frac{\sigma}{2} \right ) \, \ln \left ( \frac{\sigma + 2}{2} \right )
- \ln \left (\sigma + 1  \right ) \, \ln \left (\sigma + 2  \right )
+ \frac{2\,(\sigma+1)^2}{(\sigma+2)^3}\,\ln \left (\sigma+1 \right ) \nonumber \\
&& - \,  \frac{7\,\sigma^3+50\,\sigma^2+100\,\sigma+64}{4\,(\sigma+2)^3}\,
\ln \left ( \frac{\sigma}{2} \right )
- {1 \over 2} \, \ln \left ( \frac{\sigma + 2}{2} \right )
- {11\,\sigma^2+28\,\sigma+24 \over 8\,(\sigma+2)^2}  \bigg \}.
\hspace{0.8 cm}
\end{eqnarray}
This together with the one-loop  expression of the double spectral function $\rho^{\rm tw2}_{1}(s_1, s_2)$
constitutes one of the major technical results of this paper.
We further verify that the obtained LCSR (\ref{final result of the NLL LCSR at twist 2})
for the twist-two resolved photon correction to the magnetic coupling $g_{M^{\ast} M \gamma}$
is independent of the factorization scale $\mu$ and the renormalization scale $\nu$ of the local QCD current
$J_5 = \bar Q \, \gamma_5 \, q$  with the asymptotic photon LCDA
at one loop.

\section{The higher twist LCSR for the resolved photon effects}
\label{section: higher-twist hadronic photon correction}

The objective of this section is to compute the higher twist corrections to the magnetic coupling
from  the two-particle and three-particle photon distribution amplitudes at tree level,
up to  the twist-four accuracy. In analogy to the derivation of the twist-two LCSR
for the hadronic photon correction detailed in Section \ref{section: leading-twist hadronic photon effect},
we will need to demonstrate the hard-collinear factorization formulae of the vacuum-to-photon
correlation function (\ref{defnition: correlation function}) for the subleading twist contributions
by employing the OPE technique and the background field formalism
with the default power counting scheme (\ref{power counting scheme}).

\subsection{Two-particle higher twist corrections}

The two-particle higher twist corrections to the correlation function (\ref{defnition: correlation function})
can be obtained by keeping the subleading power terms in the light-cone expansion of the heavy-quark propagator
\begin{eqnarray}
\Pi_\mu(p,\,q)  \supset  \int d^4x \, \int \frac{d^4k}{(2\pi)^4} \,
 \frac{e^{-i(p+q+k) \cdot x}}{k^2-m_Q^2} \langle \gamma(p, \eta^{\ast}) | \bar{q}(x)\,
\left ( \sigma_{\mu \nu} \, k^\nu  + i \, m_Q\,\gamma_{\mu\perp} \right ) \, \gamma_5 \,\, q(0) | 0 \rangle \,.
\end{eqnarray}
According to the definitions of the twist-three and twist-four photon LCDA presented
in Appendix \ref{appendix: Photon distribution amplitudes}, the corresponding hard-collinear
factorization formula can be written as
\begin{eqnarray}
\Pi_\mu(p,\,q)  \supset   {e_q \, g_{\rm em} \over 4} \, \epsilon_{\mu p q \eta^{\ast}} \,
\int_0^1 d u \, \left \{  {K_2^{\rm 2PHT}((p+q)^2, q^2, u) \over [ \bar u \, (p+q)^2 + u \, q^2 - m_Q^2]^2}
+ {K_3^{\rm 2PHT}((p+q)^2, q^2, u) \over [ \bar u \, (p+q)^2 + u \, q^2 - m_Q^2]^3} \right \},
\hspace{0.5 cm}
\label{factorization formula for the two-particle higher twist effects}
\end{eqnarray}
where the two coefficient functions $K_{2\, (3)}^{\rm 2PHT}$ are given by
\begin{eqnarray}
K_2^{\rm 2PHT}((p+q)^2, q^2, u) &=& 2 \,m_Q \,f_{3\gamma}(\mu) \, \psi^{(a)}(u, \mu)
- \langle \bar{q} q \rangle(\mu)\, \mathbb{A}(u, \mu)\,,  \nonumber \\
K_3^{\rm 2PHT}((p+q)^2, q^2, u) &=& 2\, m_Q^2\, \langle \bar{q}q \rangle(\mu)\, \mathbb{A}(u,\mu) \,.
\end{eqnarray}
The dispersion representation of the tree-level factorization formula
(\ref{factorization formula for the two-particle higher twist effects})
can be further derived with the general expression for the double spectral densities
of the invariant amplitudes
\begin{eqnarray}
&& {1 \over \pi^2} \, {\rm Im}_{s_1} \, {\rm Im}_{s_2}  \,
\int_0^1 d u \, {\varphi(u, \mu)  \over [\bar u \, s_1 + u \, s_2 - m_Q^2 + i \, 0]^{n}}  \nonumber \\
&& = {1 \over \Gamma(n)} \, {d^{n - 1} \over (d \, m_Q^2)^{n -1}}  \,
\sum_k \, {(-1)^{k+1} \, c_k(\mu)\over  \Gamma(k+1) } \, (s_1 - m_Q^2)^k  \, \delta^{(k)}(s_1-s_2) \,,
\label{dispersion form of the master integral}
\end{eqnarray}
where we have employed the Taylor expansion of  $\varphi(u, \mu)$ at $u=0$
\begin{eqnarray}
\varphi(u, \mu) = \sum_k \, c_k(\mu) \, u^k \,.
\end{eqnarray}
Confronting the obtained spectral representation of (\ref{factorization formula for the two-particle higher twist effects})
with  the double hadronic dispersion relation (\ref{hadronic dispersion relation}), implementing the continuum subtraction
with the parton-hadron duality approximation and performing the double Borel transformation $(p+q)^2 \to M^2$,
$q^2 \to M^2$, we derive the LCSR for the two-particle higher twist corrections to the magnetic coupling $g_{M^{\ast} M \gamma}$
\begin{eqnarray}
&& f_P \, f_V \, \mu_P \,m_V \,   g_{M^{\ast} M \gamma}^{(\rm 2PHT)} \,
{\rm exp} \left [- \left ( {m_V^2 + m_P^2 \over M^2}  \right ) \right ]  \nonumber \\
&& = {e_q \over 2} \, \left [ m_Q \, f_{3 \gamma}(\mu) \, \psi^{(a)} \left ({1 \over 2}, \mu \right )
+ \left (  {m_Q^2 \over M^2} + {1 \over 2}  \right ) \, \langle \bar{q}q \rangle(\mu)\,  \,
\mathbb{A} \left ({1 \over 2}, \mu \right )  \right ]  \,
{\rm exp} \left ( - {2 \, m_Q^2 \over M^2}  \right ) \,.
\label{two-particle higher-twist LCSR}
\end{eqnarray}
As can be understood from  (\ref{dispersion form of the master integral}),
the obtained subleading twist sum rules (\ref{two-particle higher-twist LCSR}) at tree level
are also independent of the specific shape of the duality boundary
due to the vanishing spectral function  at $s_1 \neq s_2$.
Several  remarks on the newly derived LCSR  (\ref{two-particle higher-twist LCSR})  are in order.

\begin{itemize}

\item{Based upon the canonical behaviour of the two-particle photon LCDA
$\psi^{(a)}$ and  $\mathbb{A}$ as well as the power counting scheme for the sum rule parameters
displayed in (\ref{power counting scheme of parameters}),
we can derive the  scaling law of each individual term  appearing in   (\ref{two-particle higher-twist LCSR})
\begin{eqnarray}
g_{M^{\ast} M \gamma}^{(\rm 2PHT)} \, \big |_{\psi^{(a)}} \sim {\cal O} \left ({1 \over \Lambda_{\rm QCD}} \right)  \,,
\qquad g_{M^{\ast} M \gamma}^{(\rm 2PHT)} \, \big |_{\mathbb{A}} \sim {\cal O} \left ({1 \over \omega_0} \right)  \,.
\end{eqnarray}
In comparison with the scaling behaviour of the twist-two hadronic photon contribution
displayed in (\ref{scaling law of the twist-2 effect}),  the two-particle twist-three and twist-four corrections
to the magnetic coupling $M^{\ast} M \gamma$ are suppressed
by one and two power(s) of $\Lambda_{\rm QCD} / \omega_0$, respectively.}

\item{Applying the vector meson dominance (VMD) model,
the nonperturbative parameter $f_{3 \gamma}$ characterizing the twist-three contribution
can be expressed in terms of  the relevant $\rho$-meson couplings  \cite{Ball:2002ps}
\begin{eqnarray}
f_{3 \gamma}^{\rm VMD} \simeq - f_{\rho}^2 \, \zeta_{3, \, \rho} \,, \qquad
\zeta_{3, \, \rho}(1\, \rm GeV) = 0.10 \pm 0.05 \,.
\label{VMD for f3gamma}
\end{eqnarray}
The appearance of the strong suppression factor $\zeta_{3, \, \rho}$ in (\ref{VMD for f3gamma})
implies that the two-particle higher twist contributions to the coupling $g_{M^{\ast} M \gamma}$
will be numerically dominated by the twist-four term entering (\ref{two-particle higher-twist LCSR}).}

\end{itemize}

\subsection{Three-particle higher twist corrections}

We turn to compute the subleading power corrections to the heavy meson magnetic coupling
from the three-particle higher twist photon distribution amplitudes by employing the LCSR technique.
To achieve this goal, we  derive the corresponding  higher twist factorization formula
of the vacuum-to-photon correlation function (\ref{defnition: correlation function})
with the aid of the light-cone expansion for the quark propagator
in the background gluon/photon field \cite{Balitsky:1987bk,Belyaev:1994zk}
\begin{eqnarray}
\langle 0 | {\rm T} \, \{\bar q (x), q(0) \} | 0\rangle
 \supset   i \, g_s  \int_0^{\infty} \,\, {d^4 k \over (2 \pi)^4} \, e^{- i \, k \cdot x} \,
\int_0^1 \, d u \, \left  [ {u \, x_{\mu} \, \gamma_{\nu} \over k^2 - m_q^2}
 - \frac{(\not \! k + m_q) \, \sigma_{\mu \nu}}{2 \, (k^2 - m_q^2)^2}  \right ]
\, G^{\mu \nu}(u \, x) \nonumber \\
+ \,  i \, e_q \, g_{\rm em}  \int_0^{\infty} \,\, {d^4 k \over (2 \pi)^4} \, e^{- i \, k \cdot x} \,
\int_0^1 \, d u \, \left  [ {u \, x_{\mu} \, \gamma_{\nu} \over k^2 - m_q^2}
 - \frac{(\not \! k + m_q) \, \sigma_{\mu \nu}}{2 \, (k^2 - m_q^2)^2}  \right ]
\, F^{\mu \nu}(u \, x)\,. &&
\hspace{0.5 cm}
\end{eqnarray}
Computing the  tree-level diagrams displayed in figure \ref{fig: 3PHT at tree level}
and making use of the definitions for the three-particle photon LCDA
collected in Appendix \ref{appendix: Photon distribution amplitudes},
we find
\begin{eqnarray}
\Pi_{\mu}(p,\,q) & \supset & g_{\rm em} \, \langle \bar{q} q \rangle (\mu) \,
\epsilon_{\mu p q \eta^{\ast}} \, \int_0^1 \, du  \, \int [{\cal D} \alpha_i] \,\,
\frac{e_q \, K_{2, \, G}^{\rm 3PHT}(\alpha_i, u) + e_Q \, K_{2, \, \gamma}^{\rm 3PHT}(\alpha_i, u)}
{\left [\bar{\alpha}_u \, (p+q)^2 + \alpha_u \, q^2 - m_Q^2 + i \, 0 \right ]^2} \,,
\label{factorization formula of 3PHT at LO}
\end{eqnarray}
where we have introduced the following short-handed notations
\begin{eqnarray}
\alpha_u = \alpha_q + u \, \alpha_g, \, \qquad
\bar{\alpha}_u = 1 - \alpha_u\,,
\end{eqnarray}
and the integration measure is given by
\begin{eqnarray}
\int [{\cal D} \alpha_i] = \int_0^1 \,  d \alpha_q \, \int_0^1 \, d \alpha_{\bar{q}} \,
\int_0^1 \,  d \alpha_g \,\, \delta \left (1 - \alpha_q - \alpha_{\bar{q}} - \alpha_g \right )\,.
\end{eqnarray}
The two invariant functions $K_{2, \, G(\gamma)}^{\rm 3PHT}$ can be  expressed in terms of
the three-particle twist-four  distribution amplitudes
\begin{eqnarray}
K_{2, \, G}^{\rm 3PHT} &=& S(\alpha_i,\mu) + T_1(\alpha_i,\mu) - T_2(\alpha_i,\mu)
+ (2\,u -1)\, \left [ \tilde{S}(\alpha_i,\mu) - T_3(\alpha_i,\mu)+  T_4(\alpha_i,\mu) \right ]\,,  \nonumber \\
K_{2, \, \gamma}^{\rm 3PHT} &=&  S_{\gamma}(\alpha_i,\mu) + (2\,u -1)\,  T_4^{\gamma}(\alpha_i,\mu) \,.
\end{eqnarray}

\begin{figure}
\begin{center}
\includegraphics[width=0.70 \columnwidth]{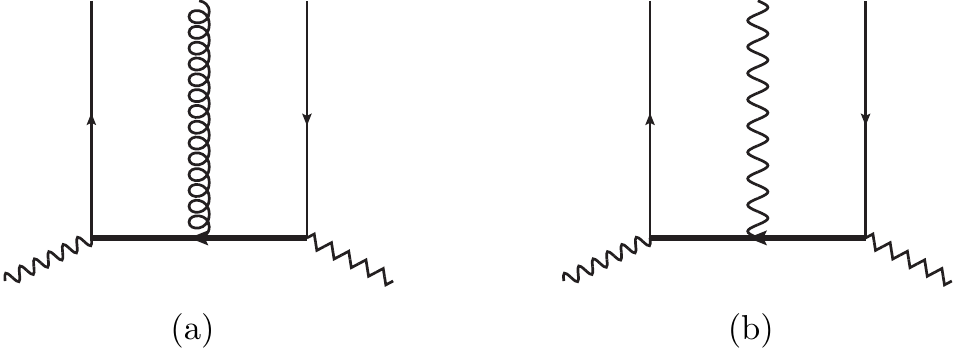}
\vspace*{0.1cm}
\caption{Diagrammatic representation of the three-particle higher twist corrections to the
vacuum-to-photon correlation function (\ref{defnition: correlation function}) at tree level.}
\label{fig: 3PHT at tree level}
\end{center}
\end{figure}

The spectral representation of the hard-collinear factorization formula
(\ref{factorization formula of 3PHT at LO}) can be constructed with
the master formula for computing the double spectral density
\begin{eqnarray}
&& {1 \over \pi^2} \, {\rm Im}_{s_1} \, {\rm Im}_{s_2}  \,
\int_0^1 d u \, \int [{\cal D} \alpha_i] \,\,
\frac{u^m \, \Phi(\alpha_i, \mu)} {\left [\bar{\alpha}_u \, s_1 + \alpha_u \, s_2 - m_Q^2 + i \, 0 \right ]^n}
 \nonumber \\
&& = {1 \over \Gamma(n)} \, {d^{n - 1} \over (d \, m_Q^2)^{n -1}}  \,
\sum_k \, {(-1)^{k+1} \, h_{m, k}(\mu)\over  \Gamma(k+1) } \, (s_1 - m_Q^2)^k  \, \delta^{(k)}(s_1-s_2) \,,
\end{eqnarray}
where the renormalization-scale dependent coefficient $h_{m, k}(\mu)$ arises from the series expansion
of  the ``effective"  photon distribution amplitude $\widehat{\Phi} (\alpha_u, m, \mu)$ defined by
\begin{eqnarray}
\widehat{\Phi}(\alpha_u, m, \mu) = \int_0^{\alpha_u} \, d \alpha_q \, \int_{\alpha_u-\alpha_q}^{1-\alpha_q} \, d \alpha_g \,
{ (\alpha_u - \alpha_q)^m \over\alpha_g^{m+1} } \, \Phi(\alpha_q, 1 - \alpha_q - \alpha_{g}, \alpha_g, \mu) \,.
\end{eqnarray}
More explicitly,
\begin{eqnarray}
\widehat{\Phi}(\alpha_u, m, \mu) = \sum_k \, h_{m, k}(\mu) \, \alpha_u^k \,.
\end{eqnarray}
We can then readily derive the tree-level LCSR for the three-particle higher twist corrections
to the  magnetic coupling $M^{\ast} M \gamma$ of our interest
\begin{eqnarray}
&& f_P \, f_V \, \mu_P \,m_V \,   g_{M^{\ast} M \gamma}^{(\rm 3PHT)} \,
{\rm exp} \left [- \left ( {m_V^2 + m_P^2 \over M^2}  \right ) \right ]  \nonumber \\
&& = - \langle \bar{q} q \rangle (\mu) \,
\bigg  \{ e_q \,  \bigg [ \widehat{S} \left ({1 \over 2}, 0, \mu \right )
+ \widehat{T}_1 \left ({1 \over 2}, 0, \mu \right )
- \widehat{T}_2 \left ({1 \over 2}, 0, \mu \right )
- \widehat{\tilde{S}} \left ({1 \over 2}, 0, \mu \right )
+ \widehat{T}_3 \left ({1 \over 2}, 0, \mu \right ) \nonumber \\
&& \hspace{3.0 cm}  - \, \widehat{T}_4 \left ({1 \over 2}, 0, \mu \right )
+ 2 \, \widehat{\tilde{S}} \left ({1 \over 2}, 1, \mu \right )
- 2 \, \widehat{T}_3 \left ({1 \over 2}, 1, \mu \right )
+ 2\,  \widehat{T}_4 \left ({1 \over 2}, 1, \mu \right )   \bigg ] \nonumber \\
&& \hspace{2.5 cm}  + \, e_Q \, \left [  \widehat{S}_{\gamma} \left ({1 \over 2}, 0, \mu \right )
 - \widehat{T}_4^{\gamma} \left ({1 \over 2}, 0, \mu \right )
 + 2 \,  \widehat{T}_4^{\gamma} \left ({1 \over 2}, 1, \mu \right )  \right ]   \bigg \}  \,
{\rm exp} \left ( - {2 \, m_Q^2 \over M^2}  \right ).
\hspace{1.0 cm}
\label{three-particle higher-twist LCSR}
\end{eqnarray}
Employing the power counting scheme (\ref{power counting scheme of parameters})
we can establish the  scaling laws for the separate three-particle twist-four contributions
from the quark-anti-quark-gluon and quark-anti-quark-photon distribution amplitudes
\begin{eqnarray}
 g_{M^{\ast} M \gamma}^{(\rm 3PHT)} \big |_{\bar q \, q \, G} \sim
 g_{M^{\ast} M \gamma}^{(\rm 3PHT)} \big |_{\bar q \, q \, \gamma}
 \sim {\cal O} \left ( {1 \over m_Q} \right ) \,,
 \end{eqnarray}
which suffer from the double suppression factor of $\Lambda_{\rm QCD}^2 / (m_Q\, \omega_0)$ compared with
the twist-two resolved photon contribution shown in (\ref{scaling law of the twist-2 effect}).
Consequently,  the three-particle higher twist corrections to the radiative bottom-meson decay form factors
are expected to be of minor importance numerically.

Putting all the different pieces together, the final expression for the heavy meson magnetic coupling
computed from the LCSR  method with the collinear photon distribution amplitudes can be written as
\begin{eqnarray}
g_{M^{\ast} M \gamma} = g_{M^{\ast} M \gamma}^{(\rm per)} + g_{M^{\ast} M \gamma}^{(\rm tw2, \, NLL)}
+ g_{M^{\ast} M \gamma}^{(\rm 2PHT)} + g_{M^{\ast} M \gamma}^{(\rm 3PHT)} \,,
\label{final LCSR for the magnetic coupling}
\end{eqnarray}
where the analytical expressions of the individual terms appearing in the right-hand
side of (\ref{final LCSR for the magnetic coupling}) are already displayed in
(\ref{sum rules for perturbative contributions}), (\ref{final result of the NLL LCSR at twist 2}),
(\ref{two-particle higher-twist LCSR}) and (\ref{three-particle higher-twist LCSR}), respectively.
We make several comments on  the higher-order corrections to the   $M^{\ast} M \gamma$ couplings
accomplished in this work.

\begin{itemize}

\item{It is apparent that constructing the desired LCSR  for the heavy meson magnetic coupling
from the vacuum-to-photon correlation function (\ref{defnition: correlation function}) is achieved
with the parton-hadron duality approximation for both the vector and pseudoscalar heavy-meson channels.
The resulting LCSR predictions for the  $M^{\ast} M \gamma$  couplings are therefore less accurate than
the ones for the semileptonic heavy-to-light transition form factors
(see, for instance \cite{Khodjamirian:2011ub,Wang:2015vgv}  and references therein).}

\item{In order to verify whether the higher-order radiative corrections can alleviate the double suppression
mechanism for the three-particle twist-four contributions at tree level,
it would be interesting to derive the NLO LCSR for the subleading twist hadronic photon contributions
to the  magnetic couplings for future development.}

\end{itemize}

\section{Numerical analysis}
\label{section: numerical analysis}

Having at our disposal the improved LCSR for the heavy meson magnetic coupling,
we are now ready to explore the phenomenological implications of the newly obtained
power corrections from the two-particle and three-particle
photon distribution amplitudes at  twist-four.
We further perform an exploratory comparison of our results for
the $M^{\ast}  M \, \gamma$ couplings with other estimates based upon
different QCD techniques and with the available experimental measurements from
the CLEO, BESIII, and BaBar Collaborations.

\subsection{Theory inputs}

\begin{table}
\centering
\renewcommand{\arraystretch}{1.3}
\resizebox{\columnwidth}{!}{
\begin{tabular}{|lll||lll|}
\hline
\hline
  Parameter
& Value
& Ref.
&  Parameter
& Value
& Ref.
\\
\hline
  $\overline{m}_u(2 \, {\rm GeV})$             & $2.32 \pm 0.10$ MeV                   & \cite{Tanabashi:2018oca} 
& $\overline{m}_c(\overline{m}_c)$             & $1.288 \pm 0.020$  GeV                & \cite{Dehnadi:2015fra} 
\\
  $\overline{m}_d(2 \, {\rm GeV})$             & $4.71 \pm 0.09$ MeV                   & \cite{Tanabashi:2018oca} 
& $\overline{m}_b(\overline{m}_b)$             & $4.193^{+0.022}_{-0.035}$  GeV        & \cite{Beneke:2014pta} 
\\
  $\overline{m}_s(2 \, {\rm GeV})$             & $92.9 \pm 0.7$  MeV                   & \cite{Tanabashi:2018oca} 
& &  &
\\
\hline
  $m_{D^{\pm}}$                                & $1869.65 \pm 0.05$ MeV                    & \cite{Tanabashi:2018oca} 
& $m_{B^{\pm}}$                                & $5279.33 \pm 0.13$ MeV                    & \cite{Tanabashi:2018oca} 
\\
  $m_{D^{0}}$                                  & $1864.83 \pm 0.05$ MeV                    & \cite{Tanabashi:2018oca} 
& $m_{B^{0}}$                                  & $5279.64 \pm 0.13$ MeV                    & \cite{Tanabashi:2018oca} 
\\
  $m_{D_s}$                                    & $1968.34 \pm 0.07$ MeV                    & \cite{Tanabashi:2018oca} 
& $m_{B_s}$                                    & $5366.88 \pm 0.17$ MeV                    & \cite{Tanabashi:2018oca} 
\\
  $m_{D^{\ast \pm}}$                           & $2010.26 \pm 0.05$ MeV                    & \cite{Tanabashi:2018oca} 
& $m_{B^{\ast \pm}}$                           & $5324.70 \pm 0.22$ MeV                    & \cite{Tanabashi:2018oca} 
\\
  $m_{D^{\ast 0}}$                             & $2006.85 \pm 0.05 $ MeV                   & \cite{Tanabashi:2018oca} 
& $m_{B^{\ast 0}}$                             & $5324.86 \pm 0.22$ MeV                    & \cite{Tanabashi:2018oca} 
\\
  $m_{D_s^{\ast}}$                              & $2112.2 \pm 0.4$ MeV                     & \cite{Tanabashi:2018oca} 
& $m_{B_s^{\ast}}$                              & $5415.4^{+1.8}_{-1.5}$ MeV               & \cite{Tanabashi:2018oca} 
\\
\hline
  $f_D|_{N_f = 2+1+1}$                        & $212.0 \pm 0.7$ MeV                    & \cite{Aoki:2019cca} 
& $f_B|_{N_f = 2+1+1}$                        & $190.0 \pm 1.3$ MeV                    & \cite{Aoki:2019cca} 
\\
  $f_{D_s}|_{N_f = 2+1+1}$                    & $249.9 \pm 0.5$ MeV                    & \cite{Aoki:2019cca} 
& $f_{B_s}|_{N_f = 2+1+1}$                    & $230.3 \pm 1.3$ MeV                    & \cite{Aoki:2019cca} 
\\
  $f_{D^{\ast}}|_{N_f = 2+1+1}$             & $223.5 \pm 8.4$ MeV                      & \cite{Lubicz:2017asp} 
& $f_{B^{\ast}}|_{N_f = 2+1+1}$             & $185.9 \pm 7.2$ MeV                      & \cite{Lubicz:2017asp} 
\\
  $f_{D_s^{\ast}}|_{N_f = 2+1+1}$            & $268.8 \pm 6.6 $ MeV                    & \cite{Lubicz:2017asp}
& $f_{B_s^{\ast}}|_{N_f = 2+1+1}$            & $223.1 \pm 5.4$  MeV                    & \cite{Lubicz:2017asp} 
\\
\hline
  $\langle \bar q q \rangle \, (1 \, {\rm GeV})$
  & $-\left ( 266 \pm 21 \, {\rm MeV} \right )^3 $                                     & \cite{Aoki:2019cca} 
& $\kappa\, (1 \, {\rm GeV})$               & $0.2 \pm 0.2$                            & \cite{Ball:2002ps,Balitsky:1989ry}
\\
  $\chi \, (1 \, {\rm GeV})$                & $(3.15 \pm 0.3) \, {\rm GeV}^{-2}$       & \cite{Ball:2002ps}
& $\kappa^{+}\, (1 \, {\rm GeV})$           & $0$                                      & \cite{Ball:2002ps,Balitsky:1989ry}
\\
  $b_2 \, (1 \, {\rm GeV})$                 & $0.07 \pm 0.07$                          & \cite{Ball:2002ps}
& $\zeta_1 \, (1 \, {\rm GeV})$             & $0.4 \pm 0.4$                            &  \cite{Ball:2002ps,Balitsky:1989ry}
\\
  $f_{3 \gamma} \, (1 \, {\rm GeV})$        & $- (4 \pm 2) \times 10^{-3} \, {\rm GeV}^{2} $             & \cite{Ball:2002ps,Balitsky:1989ry}
& $\zeta_1^{+} \, (1 \, {\rm GeV})$         & $0$                                       &  \cite{Ball:2002ps,Balitsky:1989ry}
\\
  $\omega_{\gamma}^{V}\, (1 \, {\rm GeV})$  & $3.8 \pm 1.8$                             & \cite{Ball:2002ps,Balitsky:1989ry}
& $\zeta_2^{+} \, (1 \, {\rm GeV})$         & $0$                                       &  \cite{Ball:2002ps,Balitsky:1989ry}
\\
  $\omega_{\gamma}^{A}\, (1 \, {\rm GeV})$  & $-2.1 \pm 1.0$                            & \cite{Ball:2002ps}
&                                           &                                           &
\\
\hline
  $\{s_0, \, M^2 \}$                        & $\{ 6.0 \pm 0.5, \, 4.5 \pm 1.0 \} \, {\rm GeV}^2$            &  $D^{\ast} D \gamma$
& $\{s_0, \, M^2 \}$                        & $\{34.0 \pm 1.0, \, 18.0 \pm 3.0 \} \, {\rm GeV}^2$           &  $B^{\ast} B \gamma$
\\
  $\{s_0, \, M^2 \}$                        & $\{ 7.0 \pm 0.5, \, 4.5 \pm 1.0 \} \, {\rm GeV}^2$            &  $D_s^{\ast} D_s \gamma$
& $\{s_0, \, M^2 \}$                        & $\{36.0 \pm 1.0, \, 18.0 \pm 3.0 \} \, {\rm GeV}^2$           &  $B_s^{\ast} B_s \gamma$
\\
\hline
\hline
\end{tabular}
}
\renewcommand{\arraystretch}{1.0}
\caption{Numerical input values of the theory parameters employed in the improved LCSR predictions
for the $M^{\ast}  M \, \gamma$ couplings with the subleading twist corrections. }
\label{table: input parameters}
\end{table}

To perform the numerical study of the derived LCSR (\ref{final LCSR for the magnetic coupling})
for the radiative heavy meson decay form factors,
we start discussing the numerical input values of both the QCD and hadronic parameters
as collected in Table \ref{table: input parameters}.
We employ the renormalized light-quark masses in the ${\rm \overline{MS}}$ scheme
at $\mu=2 \, {\rm GeV}$ determined from the lattice QCD calculations with $N_L=4$
($N_L$ is the number of active quark flavors at the renormalization scale $\mu$)
\cite{Tanabashi:2018oca}, which are also in good agreement with
the ${\rm N^4LO}$ QCDSR predictions \cite{Jamin:2006tj} based upon the five-loop computation
of the two-point correlation function of the scalar strangeness-changing quark currents (see \cite{Chetyrkin:2005kn}
for an independent determination of the strange-quark mass from the pseudoscalar sum rules).
In addition, the ${\rm \overline{MS}}$ charm-quark mass determined from the  ${\rm N^3LO}$ relativistic
QCDSR with the vector-current correlation function \cite{Dehnadi:2015fra} and the ${\rm \overline{MS}}$  bottom-quark mass
from the non-relativistic sum rules with high moments of $6 \leq n \leq 15 $ at ${\rm N^3LO}$ \cite{Beneke:2014pta}
will be implemented in the numerical analysis.
We further adopt the lattice QCD calculations of the decay constants for the pseudoscalar heavy mesons
with $N_f=2+1+1$ dynamical flavours summarized in  the Flavour Lattice Averaging Group (FLAG) review \cite{Aoki:2019cca},
which are compatible with the three-loop QCDSR predictions \cite{Gelhausen:2013wia}
with  higher theoretical uncertainties.
The QCD decay constants of the vector heavy-light mesons are taken from the lattice determinations
with the gauge configurations produced by European Twisted Mass Collaboration (ETMC) at $N_f=2+1+1$ \cite{Lubicz:2017asp}.
It is interesting to notice that the predicted heavy-quark spin/flavour symmetry breaking effects for the decay constant ratios
$f_{D_{(s)}^{\ast}}/f_{D_{(s)}} > 1$  in the charm sector  differ from the ones $f_{B_{(s)}^{\ast}}/f_{B_{(s)}} < 1$
in the bottom sector.

The fundamental nonperturbative functions entering the sum rules (\ref{final LCSR for the magnetic coupling})
for the heavy meson  magnetic couplings are the two-particle and
three-particle collinear photon distribution amplitudes at  twist-four.
The explicit expressions of these hadronic quantities at NLO in the conformal expansion
can be found in \cite{Ball:2002ps} (see also \cite{Wang:2018wfj} for a summary).
Among the eleven nonperturbative parameters in total,
$\langle \bar q  q \rangle$, $\chi$, $f_{3 \gamma}$, $\kappa$, $\kappa^{+}$ are defined by the QCD matrix elements of
local operators without the covariant derivative acting on the quark/gluon fields.
We use the averaged $N_f=2+1+1$ determinations of the low-energy constant $\langle \bar q  q \rangle$
at the renormalization scale $2 \, {\rm GeV}$ from  the lattice QCD calculations \cite{Aoki:2019cca},
which are consistent with the interval determined from
the chiral perturbation theory (ChPT) relations \cite{Leutwyler:1996qg,Khodjamirian:2011ub}.
An updated prediction of the magnetic susceptibility of the quark condensate from the QCDSR method
including the radiative correction to the corresponding partonic spectral density
presented in \cite{Ball:2002ps} will be adopted in what follows.
We further take the values of the normalization constant $f_{3 \gamma}$,
parameterizing  the three-body local vacuum-to-photon matrix element
$\langle \gamma(p, \eta^{\ast}) | \bar q \, g_s \, \tilde{G}_{\alpha \beta } \, \gamma_{\rho}  \, \gamma_5 \, q| 0 \rangle$,
estimated from the VMD approximation, which are also confirmed by the  QCDSR calculation independently \cite{Ball:2002ps}.
In addition, $\kappa$, $\kappa^{+}$ and the six remaining nonperturbative parameters
(i.e., $b_2$, $\omega^V_{\gamma}$, $\omega^A_{\gamma}$, $\zeta_1$, $\zeta_1^{+}$, $\zeta_2^{+}$)
entering the ``P"-wave terms in the conformal expansion of the photon LCDA are taken from
the numerical evaluations obtained in \cite{Balitsky:1989ry}.
As  discussed in detail in \cite{Ball:2002ps}, the parameter $\zeta_2$ entering
the three-particle twist-four photon distribution amplitudes
can be expressed in terms of $\zeta_1$ and $\zeta_2^{+}$,
thanks to the Ferrara-Grillo-Parisi-Gatto theorem \cite{Ferrara:1972xq}.
The renormalization scale dependence of the twist-three and twist-four parameters
at the leading-logarithmic (LL) approximation can be derived by solving
the corresponding one-loop evolution equations \cite{Ball:1998sk,Ball:2002ps}.

The determinations of the Borel parameter $M^2$ and the effective threshold $s_0$
can be achieved by applying the standard criteria imposed in the LCSR calculations:
the smallness of subleading twist contributions in the OPE,
and simultaneously, the suppression of higher state contributions
(see \cite{Colangelo:2000dp} for a modern review).
The obtained intervals of theses ``internal" sum rule parameters presented
in Table \ref{table: input parameters} are
in agreement with the previous choices adopted in the LCSR analysis of
heavy-to-light decay form factors \cite{Khodjamirian:2009ys,Khodjamirian:2011ub}.
Furthermore,  we will vary the factorization scale $\mu$ appearing in the
NLL LCSR of the twist-two hadronic photon contribution (\ref{final result of the NLL LCSR at twist 2})
in the range $[1.0, 2.0] \, {\rm GeV}$ around the default value $\bar{m}_c(\bar{m}_c)$
for the magnetic  couplings $D_{q}^{\ast} D_{q} \gamma$
and in $[m_b/2, 2 \, m_b]$ around $\bar{m}_b(\bar{m}_b)$ for the counterpart bottom-meson couplings.
The  renormalization scale for the QCD pseudoscalar current will be further taken as $\nu = m_Q$
\cite{Beneke:2005gs,Gao:2019lta}.

\subsection{Theory predictions for radiative heavy meson form factors}

\begin{table}[t]
\begin{center}
\renewcommand{\arraystretch}{1.3}
\renewcommand{\multirowsetup}{\centering}
\begin{tabular}{|c||c|c|c||c|c|c|}
\hline
\hline
& $D^{\ast +} \, D^{+} \, \gamma$ & $D^{\ast 0} \, D^{0} \, \gamma$ & $D_s^{\ast +} \,  D_s^{+} \, \gamma$
& $B^{\ast +} \, B^{+} \, \gamma$ & $B^{\ast 0} \, B^{0} \, \gamma$ &  $B_s^{\ast 0} \, B_s^{0} \, \gamma$  \\
\hline
$g_{M^{\ast} M \gamma}^{(\rm per)} \, ({\rm GeV}^{-1})$  & $-0.032$  & $0.73$ & $0.024$ & $0.84$ & $-0.56$ & $-0.47$ \\
\hline
$g_{M^{\ast} M \gamma}^{(\rm tw2, \, LL)} \, ({\rm GeV}^{-1})$  & $-0.50$  & $0.99$ & $-0.41$ & $0.96$ & $-0.48$ & $-0.37$ \\
\hline
$g_{M^{\ast} M \gamma}^{(\rm tw2, \, NLL)} \, ({\rm GeV}^{-1})$  & $-0.45$  & $0.89$ & $-0.36$ & $0.79$ & $-0.40$ & $-0.31$ \\
\hline
$g_{M^{\ast} M \gamma}^{(\rm 2PHT)} \, ({\rm GeV}^{-1})$  & $0.18$  & $-0.37$ & $0.14$ & $-0.18$ & $0.089$ & $0.067$ \\
\hline
$g_{M^{\ast} M \gamma}^{(\rm 3PHT)} \, ({\rm GeV}^{-1})$  & $0.14$  & $0.23$ & $0.11$ & $-0.017$ & $-0.036$ & $-0.027$ \\
\hline
$g_{M^{\ast} M \gamma} \, ({\rm GeV}^{-1})$ & $-0.15$  & $1.48$ & $-0.079$ & $1.44$ & $-0.91$ & $-0.74$ \\
\hline
\hline
\end{tabular}
\end{center}
\caption{The numerical results of all separate terms entering the obtained LCSR (\ref{final LCSR for the magnetic coupling})
for the $M^{\ast} M \gamma$ couplings  with the central values of theory inputs.  }
\label{pattern for the magnetic coupling}
\end{table}

We are now in a position to investigate the numerical impacts of the perturbative QCD corrections
and the subleading twist corrections to the resolved photon contributions for the heavy meson magnetic couplings.
It is apparent from  Table \ref{pattern for the magnetic coupling} that the twist-two hadronic photon
corrections to the bottom-meson couplings $B_q^{\ast} \, B_q \, \gamma$  are numerically comparable to
the corresponding ``point-like" photon contributions and the  QCD radiative corrections will reduce the tree-level
twist-two predictions by approximately an amount of $20 \, \%$.
In addition, such long-distance twist-two  corrections turn out to be
the most significant contributions for the charm-meson magnetic couplings,
particularly for  the radiative $D^{+ \ast} \to  D^{+} \, \gamma$ and  $D_s^{+ \ast} \to D_s^{+} \, \gamma$
decay form factors, where the leading power ``point-like" photon contributions are  numerically  suppressed
due to the strong cancellation between  the photon radiation off the charm and light quarks as already mentioned
in Section \ref{section: theory summary}.
The SU(3)-flavour symmetry breaking effect between the two magnetic couplings
$B^{0\ast} \, B^{0} \, \gamma$ and  $B_s^{0 \ast} \, B_s^{0} \, \gamma$ is evaluated
to be approximately ${\cal O} (20 \, \%)$ based upon the LCSR technique,
in analogy to the ones for the semileptonic $B$-meson form factors \cite{Lu:2018cfc,Gao:2019lta}.
Furthermore,  our predictions for the two-particle and three-particle higher twist corrections to the
$B_q^{\ast} \, B_q\, \gamma$ couplings at tree level imply that the subleading twist terms appearing
in the bottom-meson sum rules are indeed of minor phenomenological importance, in contrary to the observed
patterns for the counterpart charm-meson couplings as displayed in  Table \ref{pattern for the magnetic coupling}.

We proceed to present the individual uncertainties for the predicted heavy meson magnetic couplings
due to variations of  the input parameters
in Table \ref{table for the uncertainties of the magnetic couplings},
where the total theory  uncertainties obtained by adding all separate uncertainties in quadrature
are further shown for completeness.
Evidently,  the dominant theory uncertainties of the LCSR predictions for the bottom-meson
couplings arise from the variations of the quark condensate density $\langle \bar{q} q \rangle$
as well as its  magnetic susceptibility $\chi$.
As expected, the LCSR predictions for the charm-meson magnetic couplings suffer from larger theory uncertainties
than the ones for the corresponding bottom-meson couplings.
It needs to be pointed out further  that the  substantial uncertainties for the two couplings $D^{\ast +}  D^+  \gamma$
and $D_s^{\ast +}D_s^{+} \gamma$ can be attributed to the  almost complete cancellation of the
two different pieces  entering the ``point-like" photon contribution (\ref{sum rules for perturbative contributions}),
which are classified in terms of the electric-charge factors $e_q$ and $e_Q$ respectively.

\begin{table}[t]
\begin{center}
\renewcommand{\arraystretch}{1.3}
\renewcommand{\multirowsetup}{\centering}
\begin{tabular}{|c||c|c|c||c|c|c|}
\hline
\hline
  & $g_{D^{\ast +}  D^+  \gamma}$ & $g_{D^{\ast 0} D^{0} \gamma}$  &  $g_{D_s^{\ast +}D_s^{+} \gamma}$
  & $g_{B^{\ast +} B^{+} \gamma}$ & $g_{B^{\ast 0} B^{0} \gamma}$  &  $g_{B_s^{\ast 0} B_s^{0} \gamma}$  \\
& $({\rm GeV}^{-1})$  & $({\rm GeV}^{-1})$  &  $({\rm GeV}^{-1})$
&  $({\rm GeV}^{-1})$ & $({\rm GeV}^{-1})$  &  $({\rm GeV}^{-1})$  \\
\hline
central value  & $-0.15$  & $1.48$ & $-0.079$
& $1.44$ & $-0.91$ & $-0.74$ \\
\hline
$\Delta f_{P}$ & $\pm 0.00$   & $\pm 0.02$ & $\pm 0.001$& $\pm 0.03$& $\pm 0.02$ & $\pm 0.01$  \\
\hline
$\Delta f_{V}$ & $\pm 0.01$& $\pm 0.05$ & $\pm 0.002$& $\pm  0.05$  & $\pm0.03$& $\pm 0.02$  \\
\hline
$\Delta m_{Q}$ & $\pm 0.00$ & $\pm 0.02$ & $^{+0.005}_{-0.004}$ & $^{+0.06}_{-0.04}$
& $^{+0.02}_{-0.04}$ & $^{+0.02}_{-0.03}$  \\
\hline
$\Delta s_{0}$ & $\pm 0.00$ & $^{+0.07}_{-0.09}$ & $\pm 0.004$  & $^{+0.05}_{-0.06}$
& $\pm 0.04$ &$^{+0.03}_{-0.02}$  \\
\hline
$\Delta M^2$ & $^{+0.07}_{-0.03}$ & $\pm 0.01$  & $^{+0.050}_{-0.019}$
& $\pm 0.02$ & $^{+0.01}_{-0.02}$ & $\pm 0.02$  \\
\hline
$\Delta \mu$ & $^{+0.00}_{-0.02}$ & $^{+0.05}_{-0.00}$  & $^{+0.001}_{-0.019}$ & $^{+0.00}_{-0.04}$
& $^{+0.02}_{-0.00}$ & $^{+0.02}_{-0.00}$  \\
\hline
$\Delta \langle \bar{q} q \rangle$ & $^{+0.04}_{-0.05}$ & $^{+0.22}_{-0.19}$ &$^{+0.033}_{-0.039}$
& $^{+0.17}_{-0.14}$ & $^{+0.08}_{-0.09}$ &$^{+0.06}_{-0.07}$  \\
\hline
$\Delta \chi$ & $\pm 0.04$ & $\pm 0.08$  & $\pm 0.034$& $\pm 0.08$ & $\pm 0.04$& $\pm 0.03$ \\
\hline
$\Delta b_2$ & $\pm 0.05$ & $\pm 0.10$  & $\pm 0.042$& $\pm 0.07$ & $\pm 0.04$& $\pm 0.03$ \\
\hline
$\Delta f_{3 \gamma}$ & $\pm 0.03$ & $\pm 0.06$  & $\pm 0.024$& $\pm 0.03$ & $\pm 0.01$& $\pm 0.03$  \\
\hline
$\Delta \omega_{\gamma}^{V}$ & $\pm 0.02$ & $\pm 0.03$  & $\pm 0.014$& $\pm 0.01$ & $\pm 0.01$& $\pm 0.01$ \\
\hline
$\Delta \kappa$ & $\pm 0.01$ & $\pm 0.03$  & $\pm 0.011$& $\pm 0.01$ & $\pm 0.01$& $\pm 0.00$  \\
\hline
$\Delta_{\rm tot}$ & $_{-0.10}^{+0.11}$  & $_{-0.27}^{+0.29}$ & $_{-0.078}^{+0.086}$
& $_{-0.20}^{+0.22}$ & $_{-0.13}^{+0.12}$ & $_{-0.10}^{+0.09}$  \\
\hline
\hline
\end{tabular}
\end{center}
\caption{Summary of the individual theory uncertainties for the heavy meson magnetic couplings
$M^{\ast} M \gamma$ predicted from the LCSR (\ref{final LCSR for the magnetic coupling}).
The negligibly small uncertainties due to variations of the remaining input parameters are not
displayed  here explicitly, but are already taken into account in the determinations
of the total errors $\Delta_{\rm tot}$. }
\label{table for the uncertainties of the magnetic couplings}
\end{table}

Finally, we compare our predictions of the heavy meson magnetic couplings with
the previous theory determinations obtained from different QCD techniques and phenomenological models
in Table \ref{table for the comparision of the magnetic couplings},
where the HH$\chi$PT results including both the ${\cal O}(m_q^{1/2})$ corrections to the photon
radiation off the light quarks and the subleading power corrections to the photon
coupling to the heavy quarks are derived from the following formula \cite{Amundson:1992yp}
\begin{eqnarray}
g_{M^{\ast} M \gamma}
= {e_Q \over m_Q} \, \left (  1 + {2 \over 3}\, {\bar \Lambda \over m_Q} \right )
+ e_q \, \beta + \delta \mu^{(\ell)}_{q} \,.
\label{ChPT formulae}
\end{eqnarray}
The non-perturbative HQET parameter $\bar \Lambda$ characterizing the canonical size
of the power corrections to the heavy-quark mass limit is estimated to be
$\bar \Lambda \simeq 0.50 \, {\rm GeV}$ from the two-point sum rule method \cite{Bagan:1991sg,Neubert:1992fk}.
The effective coupling $\beta$ appearing in the HH$\chi$PT Lagrangian \cite{Cheng:1992xi,Stewart:1998ke}
has been  determined to be $\beta= (3.41 \pm 0.16) \, {\rm GeV}^{-1}$ \cite{Grinstein:2015aua},
consistent with the predictions from the nonrelativistic quark model \cite{Amundson:1992yp}.
The explicit expressions of the SU(3)-flavour symmetry breaking terms $\delta \mu^{(\ell)}_{q}$
generated by the pion and kaon loops are given by  \cite{Amundson:1992yp}
\begin{eqnarray}
\delta \mu^{(\ell)}_{u} = - {g_{\pi}^2 m_K \over 4 \, \pi \, f_K^2}
- {g_{\pi}^2 m_{\pi} \over 4 \, \pi \, f_{\pi}^2} \,,
\qquad
\delta \mu^{(\ell)}_{d} = - {g_{\pi}^2 m_{\pi} \over 4 \, \pi \, f_{\pi}^2}  \,,
\qquad
\delta \mu^{(\ell)}_{s} =
- {g_{\pi}^2 m_K \over 4 \, \pi \, f_K^2}  \,.
\end{eqnarray}
We will employ the interval of the strong coupling $g_{\pi}=0.57 \pm 0.01$
extracted from the experimental measurements for the  decay width
of $D^{\ast +} \to D^{0} \pi^{+}$ \cite{Lees:2013uxa,Lees:2013zna}.
Apparently,  our LCSR determinations of the heavy meson magnetic couplings
are in reasonable agreement with the achieved HH$\chi$PT predictions
within the theory uncertainties  shown
in Table \ref{table for the comparision of the magnetic couplings}.

\begin{table}[t]
\begin{center}
\renewcommand{\arraystretch}{1.5}
\renewcommand{\multirowsetup}{\centering}
\resizebox{\columnwidth}{!}{
\begin{tabular}{|c||c|c|c||c|c|c|}
\hline
\hline
  & $g_{D^{\ast +}  D^+  \gamma}$ & $g_{D^{\ast 0} D^{0} \gamma}$  &  $g_{D_s^{\ast +}D_s^{+} \gamma}$
  & $g_{B^{\ast +} B^{+} \gamma}$ & $g_{B^{\ast 0} B^{0} \gamma}$  &  $g_{B_s^{\ast 0} B_s^{0} \gamma}$  \\
& $({\rm GeV}^{-1})$  & $({\rm GeV}^{-1})$  &  $({\rm GeV}^{-1})$
&  $({\rm GeV}^{-1})$ & $({\rm GeV}^{-1})$  &  $({\rm GeV}^{-1})$  \\
\hline
this work  & $-0.15^{+0.11}_{-0.10}$  & $1.48^{+0.29}_{-0.27}$ & $-0.079^{+0.086}_{-0.078}$
& $1.44^{+0.22}_{-0.20}$ & $-0.91^{+0.12}_{-0.13}$ & $-0.74^{+0.09}_{-0.10}$ \\
\hline
HH$\chi$PT \cite{Amundson:1992yp} & $-0.27 \pm 0.05$& $2.19 \pm 0.11$ & $0.041 \pm 0.056$
& $1.45 \pm 0.11$  & $-1.01 \pm 0.05$  & $-0.70 \pm 0.06$  \\
\hline
HQET+VMD \cite{Colangelo:1993zq} & $-0.29^{+0.19}_{-0.11}$ & $1.60^{+0.35}_{-0.45}$ & $-0.19^{+0.19}_{-0.08}$
& $0.99^{+0.19}_{-0.23}$ & $-0.58^{+0.12}_{-0.10}$ & $-$  \\
\hline
HQET+CQM \cite{Cheung:2014cka} & $-0.38^{+0.05}_{-0.06}$ & $1.91 \pm 0.09$
& $-$  & $1.45^{+0.11}_{-0.12}$ & $-0.82^{+0.06}_{-0.05}$ & $-$ \\
\hline
Lattice QCD \cite{Becirevic:2009xp} & $-0.2 \pm 0.3$ & $2.0 \pm 0.6$ & $-$& $-$  & $-$ & $-$  \\
\hline
LCSR \cite{Aliev:1995zlh} & $-0.50 \pm 0.12$ & $1.52 \pm 0.25$ & $-$ & $1.68 \pm 0.17$
& $-0.85 \pm 0.17$ & $-$  \\
\hline
QCDSR \cite{Aliev:1994nq} & $-0.19^{+0.03}_{-0.02}$ & $0.62 \pm 0.03$ & $-0.20 \pm 0.03$
& $-$ & $-$ & $-$  \\
\hline
RQM \cite{Goity:2000dk} & $-0.44\pm 0.06$ & $2.15 \pm 0.11$  & $-0.19 \pm 0.03$
& $1.66 \pm 0.11$ & $-0.93 \pm 0.05$ & $0.65 \pm 0.03$  \\
\hline
experiment \cite{Bartelt:1997yu,Aubert:2005ik,Ablikim:2014mww}
& $-0.47 \pm 0.06$   & $1.77 \pm 0.03$   & $-$    & $-$   & $-$ & $-$  \\
\hline
\hline
\end{tabular}
}
\end{center}
\caption{Comparisons of the LCSR calculations for the heavy-meson magnetic couplings
with the previous determinations and the available experimental data.}
\label{table for the comparision of the magnetic couplings}
\end{table}

Instead of computing the hadronic matrix elements  of the electromagnetic currents
of the light-flavour quarks from the HH$\chi$PT technique,
the VMD approximation and the covariant quark model (CQM) are employed to
evaluate the second and third terms $ e_q \, \beta + \delta \mu^{(\ell)}_{q}$
entering the expression for the heavy meson magnetic coupling (\ref{ChPT formulae})
in  \cite{Colangelo:1993zq} and \cite{Cheung:2014cka}, respectively.
In general we observe a fair agreement of the resulting predictions from  three
different approaches, with the exceptions of $g_{D^{\ast +}  D^+  \gamma}$
and $g_{B^{\ast 0} B^{0} \gamma}$.
Moreover, the available lattice QCD results of the charm-meson  magnetic couplings
$g_{D^{\ast +}  D^{+}  \gamma}$ and $g_{D^{\ast 0}  D^{0}  \gamma}$   \cite{Becirevic:2009xp}
obtained by employing the gauge field configurations from the QCDSF Collaboration \cite{AliKhan:2003br}
are compatible with our LCSR calculations.
Confronting the previous LCSR predictions including the two-particle
higher twist corrections \cite{Aliev:1995zlh}  with
the three-point QCDSR estimates at tree level \cite{Aliev:1994nq} already reveals some tensions
of the determined intervals for $g_{D^{\ast +}  D^{+}  \gamma}$ and $g_{D^{\ast 0}  D^{0}  \gamma}$,
which could be traced back to the unaccounted subleading twist corrections from the collinear
photon distribution amplitudes in the above-mentioned LCSR and the systematic uncertainty of the classical sum rule method
due to the contamination from the non-diagonal transitions  of the ground state to
excited states \cite{Braun:1997kw}.
In addition, the predicted bottom-meson magnetic couplings from the relativistic quark model (RQM)
\cite{Goity:2000dk} are in excellent agreement with our LCSR results.
However, the aforementioned RQM predictions for the counterpart charm-meson magnetic couplings are consistently
higher  in magnitude than our determinations.
We also mention in passing that our prediction for the $B^{\ast 0} B \gamma$   coupling
is consistent with the one determined in \cite{Khodjamirian:2015dda}.

We further collect the extracted values of $g_{D^{\ast +}  D^{+}  \gamma}$ from the
CLEO data on the branching ratio of the radiative decay $D^{\ast +}  \to D^+  \gamma$ \cite{Bartelt:1997yu}
and the Particle Data Group (PDG) average of
$\Gamma(D^{\ast +}) = 83.4 \pm 1.8 \, {\rm keV}$ \cite{Tanabashi:2018oca}
in Table \ref{table for the comparision of the magnetic couplings},
where the displayed interval for the neutral charm-meson magnetic coupling
is obtained from the experimental measurements of ${\cal BR}(D^{\ast 0}  \to D^{0}  \, \gamma)$
\cite{Tanabashi:2018oca} and the estimated result of the total decay width
$\Gamma(D^{\ast 0})=55.4 \pm 1.4 \, {\rm keV}$
by applying the well-known isospin symmetry relations of
the following strong coupling constants \cite{Belyaev:1994zk}
\begin{eqnarray}
g_{D^{\ast +} D^{0} \pi^{+}} = - \sqrt{2} \,g_{D^{\ast +} D^{+} \pi^{0}}
= \sqrt{2} \,g_{D^{\ast 0} D^{0} \pi^{0}}  \,.
\end{eqnarray}
As far as the magnitude is concerned, our computations of the  two couplings
$g_{D^{\ast +}  D^{+}  \gamma}$ and $g_{D^{\ast 0}  D^{0}  \gamma}$ yield
somewhat lower  values than determined from the experimental measurements
of the corresponding radiative decay widths.
It would be interesting to investigate whether such discrepancies can be
resolved by taking into account the NLO QCD corrections to the ``point-like"
photon contributions in the LCSR framework.


\section{Conclusion}
\label{section: conclusion}

In the present paper we have computed the twist-two hadronic  photon corrections
to the radiative heavy meson decay  form factors at the NLL accuracy with the aid of the
LCSR technique.  The resummation improved hard-collinear factorization formula for the vacuum-to-photon correlation
function defined with the two interpolating currents for the vector and pseudoscalar heavy mesons
was established by applying the evanescent operator approach and the two-loop RG equation of
the twist-two photon distribution amplitude.
We  derived the double spectral representation of the resulting QCD factorization formula
and subsequently  construed the desired LCSR for the  twist-two resolved photon contributions
to the magnetic couplings $M^{\ast} M \gamma$ analytically by implementing the parton-hadron duality ansatz
and the double Borel transformation.
The subleading twist corrections from both the two-particle and three-particle photon LCDA
up to and including twist-four were further evaluated at tree level from the same LCSR method,
taking advantage of the background field formalism.
The newly determined double spectral densities for the subleading twist contributions
enabled us to  perform the continuum subtractions analytically in constructing
the higher-twist sum rules on the light-cone.

Exploring the obtained LCSR for the magnetic $M^{\ast} M \gamma$  couplings  numerically£¬
we observed that the twist-two hadronic photon corrections to the bottom-meson couplings
are comparable to the counterpart ``point-like" photon contributions despite of the
$\Lambda_{\rm QCD}/\omega_0$ suppression.
In particular, such structure-dependent hadronic corrections gave rise to
the dominant contributions to the  two magnetic couplings $D^{\ast +}  D^{+}  \gamma$
and $D_{s}^{\ast +}  D_{s}^{+}  \gamma$, confirming the previous observations concluded
from the three-point sum rule calculation \cite{Aliev:1994nq}
and the HH$\chi$PT analysis \cite{Amundson:1992yp}.
Moreover, the predicted NLL QCD corrections to the twist-two resolved photon contributions
of the bottom-meson magnetic couplings can generate approximately $20 \%$ reduction
to the corresponding tree-level determinations.
We further noticed that the subleading twist contributions to the charm-meson couplings
from the two-particle and three-particle photon distribution amplitudes turned out to be
more pronounced than  the counterpart effects for the bottom-meson magnetic couplings.
Confronting our LCSR predictions with  various  evaluations from the diverse QCD techniques
generally led to a fair agreement for the obtained values of the $M^{\ast} M \gamma$ couplings
within the theory uncertainties.

Developing the LCSR for the heavy-meson radiative decay form factors beyond the current work
can be pursued  further  in different directions.
First, computing the NLO QCD corrections to the leading power ``point-like" photon contributions
will be in high demand in order to achieve a better understanding of the observed $g_{D^{\ast 0} D^{0} \gamma}$
tension between the LCSR predictions and the CLEO measurements.
The technical challenges of constructing such NLO sum rules arise from both the two-loop computations of
the vacuum-to-photon correlation function (\ref{defnition: correlation function})
and the analytical determinations of the double spectral densities entering the dispersion representation
of the derived QCD factorization formula.
Second, it will be of both technical and conceptual interest to compute the perturbative QCD corrections
to the higher-twist contributions  at twist-four systematically in the LCSR framework.
Extracting the hard matching coefficients appearing in the higher-twist factorization formula
for the correlation function  (\ref{defnition: correlation function}) at NLO will be complicated
by the nontrivial infrared subtractions due to the renormalization mixing of
the different  light-ray collinear operators.
Third, updating the non-perturbative parameters in the conformal expansion of the photon distribution amplitudes
with the standard QCDSR approach will be also of phenomenological importance to pin down the theory uncertainties
for the LCSR calculations of the magnetic $M^{\ast} M \gamma$  couplings.

\subsection*{Acknowledgements}

C.D.L is supported in part by the National Natural Science Foundation
of China (NSFC) with Grant No. 11521505 and 11621131001.
Y.M.W acknowledges support from the National Youth Thousand Talents Program,
the Youth Hundred Academic Leaders Program of Nankai University,
the  National Natural Science Foundation of China  with
Grant No. 11675082 and 11735010, and  the Natural Science Foundation of Tianjin
with Grant No. 19JCJQJC61100.
The work of Y.B.W is supported in part by the NSFC with Grant No. 11847238.
Y.M.W also would like to thank  Martin Beneke for the warm hospitality
during his visit at Technical University Munich, Germany.


\appendix

\section{Useful one-loop integrals}
\label{appendix: Feynman integrals at one loop}

We collect in this appendix the analytical results of various one-loop Feynman integrals
for evaluating the NLO QCD corrections to the twist-two resolved photon contributions
displayed in Section \ref{section: leading-twist hadronic photon effect}.
\begin{eqnarray}
I_a &=& \int [{\cal D} \ell] \,
{m_Q^2 \over [(\ell + u \, p)^2  + i0] [(\ell - \bar u \, p - q)^2 -m_Q^2 + i0] [\ell^2 + i 0]} \nonumber \\
&=&  {1 \over r_2 - r_1} \,
\left \{ \left [ {1 \over \epsilon}  + \ln {\mu^2 \over m_Q^2} - \ln [(1-r_1) (1-r_2)] \right ] \,
\ln  {1-r_1 \over 1-r_2} + {\rm Li}_2(r_2) - {\rm Li}_2(r_1) \right \},   \\
I_{a,\alpha} &=& \int [{\cal D} \ell] \,
{m_Q^2 \, \ell_\alpha  \over [(\ell + u \, p)^2  + i0] [(\ell - \bar u \, p - q)^2 -m_Q^2 + i0] [\ell^2 + i 0]} \nonumber \\
&=&  I_{A}\,(up)_\alpha + I_{B}\,(p+q)_\alpha \,,   \\
I_{a,\alpha\beta} &=& \int [{\cal D} \ell] \,
{\ell^\perp_\alpha\, \ell^\perp_\beta  \over [(\ell + u \, p)^2  + i0]
[(\ell - \bar u \, p - q)^2 -m_Q^2 + i0] [\ell^2 + i 0]}
\nonumber \\
&=& \frac{g^\perp_{\alpha\beta}}{4} \, \left \{{1 \over \epsilon}  + \ln {\mu^2 \over m_Q^2}
+\frac{1}{r_2-r_1}\,
\left [\frac{(1-r_1)^2}{r_1}\,\ln(1-r_1) - \frac{(1-r_2)^2}{r_2}\,\ln(1-r_2) \right ]
+ 3\! \right \},  \hspace{0.8 cm}  \\
&& \nonumber \\
I_{b,\alpha} &=& \int [{\cal D} \ell] \,
{m_Q^2 \, \ell_\alpha \over [(\ell - \bar u \, p)^2  + i0]
[(\ell - \bar u \, p - q)^2 -m_Q^2 + i0] [\ell^2 + i 0]} \nonumber \\
&=&  - I_A\big|_{r_2 \to r_3}\,(\bar u p)_\alpha
+ I_B\big|_{r_2 \to r_3}\,q_\alpha \,,   \\
&& \nonumber \\
I_c &=& \int [{\cal D} \ell] \,
{1 \over  [(\ell - \bar u \, p - q)^2 -m_Q^2 + i0] [\ell^2 + i 0]} \nonumber \\
&=&  {1 \over \epsilon}  + \ln {\mu^2 \over m_Q^2}  - \left (1 -{1 \over r_1} \right ) \, \ln (1-r_1) + 2 \,, \\
I_{c,\alpha} &=& \int [{\cal D} \ell] \,
{\ell_\alpha \over  [(\ell - \bar u \, p - q)^2 -m_Q^2 + i0] [\ell^2 + i 0]} \nonumber \\
&=&  \frac{1}{2} \, \left [ {
1 \over \epsilon}  + \ln {\mu^2 \over m_Q^2}  -
\left (\frac{1-r_1}{r_1} \right )^2 \, \ln (1-r_1) + 2 -\frac{1}{r_1} \right ]\,
(\bar up+q)_\alpha \,, \\
&& \nonumber \\
I_{d,\alpha} &=& \int [{\cal D} \ell] \,
{ m_Q^2 \, \bar n\cdot \ell \, \ell_\alpha  \over [(\ell + u \, p)^2  + i0]
[(\ell - \bar u \, p)^2  + i0]
[(\ell - \bar u \, p - q)^2 -m_Q^2 + i0] [\ell^2 + i 0]} \nonumber \\
&=&  \frac{1}{n\cdot p}\, \left [ I_{b,\alpha} -I_{a,\alpha} \right ],
\end{eqnarray}
with
\begin{align}
I_A = &~ \frac{1}{r_2-r_1}\, \left [
(r_1-1)\,I_a - 2\,r_2\,I_B + I_c \right ], \\
I_B = &~ \frac{1}{r_2-r_1}\, \left [
\frac{1-r_1}{r_1}\,\ln(1-r_1) - \frac{1-r_2}{r_2}\,\ln(1-r_2) \right].
\end{align}
The integration measure is defined as
\begin{eqnarray}
\int [{\cal D} \ell] = {(4 \pi)^2 \over i} \, \left [ {\mu^2 \, e^{\gamma_E} \over 4 \pi} \right ]^{\epsilon}\,
\int {d^D \ell \over (2 \, \pi)^D},
\end{eqnarray}
and the dimensionless parameters $r_1 = (\bar u \, p+q)^2/m_Q^2$, $r_2 = (p+q)^2/m_Q^2$ and  $r_3 = q^2/m_Q^2$
are further introduced for brevity.

\section{Master formulae for the spectral representations}
\label{appendix: master formulae for double spectral densities}

Here we collect the necessary identities for computing the double spectral density appearing in
the dispersion representation of the NLL QCD factorization formula of the correlation function
(\ref{defnition: correlation function}) at the twist-two accuracy.
\begin{eqnarray}
&& {1 \over \pi} \, {\rm Im}_{r_3} \,\, \int_{r_2}^{r_3} d \rho \,
{ \ln (1-\rho) \over 1 - \rho} \, f(\rho)
= - \int_{r_2}^{r_3} d \rho \,\, \left [ {\theta(\rho-1) \over 1 - \rho} \right ]_{+}  \,  f(\rho)
+ \ln (r_3 -1) \, f(1) \,,
\label{master dispersion formula first}
\\
\nonumber \\
&& {1 \over \pi} \, {\rm Im}_{r_3} \, \int_{r_2}^{r_3} d \rho \,\,
{ {\rm Li}_2(\rho) \over 1 - \rho} \, f(\rho)
= \int_{r_2}^{r_3} d \rho \,\, \left [ {\theta(\rho-1) \, \ln \rho \over 1 - \rho} \right ]_{+}  \,  f(\rho)
+ \left [  {\rm Li}_2(1-r_3) + {\pi^2 \over 6} \right ]  \, f(1) \,, \hspace {1.0 cm}
\\
\nonumber \\
&& {1 \over \pi} \, {\rm Im}_{r_3} \, \int_{r_2}^{r_3} d \rho \,\,
{ \ln^2(1-\rho) \over 1 - \rho} \, f(\rho)
= -2 \, \int_{r_2}^{r_3} d \rho \,\,
\left [ {\theta(\rho-1) \, \ln (\rho-1) \over 1 - \rho} \right ]_{+}  \,  f(\rho) \nonumber \\
&& \hspace{6.0 cm} + \left [  \ln^2(r_3-1) - {\pi^2 \over 3} \right ]  \, f(1) \,,
\\
\nonumber \\
&& {1 \over \pi} \, {\rm Im}_{r_3} \, \int_{r_2}^{r_3} d \rho \,\,
{ \ln(1-r_3) \over 1 - \rho} \, f(\rho)
= - \, \int_{r_2}^{r_3} d \rho \,\, \left [ {\theta(1 - \rho) \over 1 - \rho}
+  {\theta(1-\rho) \over 1- \rho}  \right ]_{+}  \,  f(\rho)
\nonumber \\
&& \hspace{6.0 cm} +  \ln {(r_3-1)^2 \over 1-r_2} \, f(1) \,,
\\
\nonumber \\
&& {1 \over \pi} \, {\rm Im}_{r_3} \, \int_{r_2}^{r_3} d \rho \,\,
{ {\rm Li}_2(r_3) \over 1 - \rho} \, f(\rho)
=  \, \int_{r_2}^{r_3} d \rho \,\, \left [ {\theta(1 - \rho) \over 1 - \rho}
+  {\theta(\rho-1) \over 1 - \rho}  \right ]_{+}  \,  \ln r_3 \,\,  f(\rho)
\nonumber \\
&& \hspace{6.0 cm} - \left \{ {\rm Li}_2(1-r_3) + \ln r_3 \, \ln {(r_3-1)^2 \over 1-r_2}
- {\pi^2 \over 6} \right \}  \, f(1) \,,
\\
\nonumber \\
&& {1 \over \pi} \, {\rm Im}_{r_3} \, \int_{r_2}^{r_3} d \rho \,\,
{ \ln^2(1-r_3) \over 1 - \rho} \, f(\rho)
= - 2 \, \int_{r_2}^{r_3} d \rho \,\, \left [ {\theta(1 - \rho) \over 1 - \rho}
+  {\theta(\rho-1) \over 1 - \rho}  \right ]_{+}  \,  \ln (r_3-1) \,\,  f(\rho)
\nonumber \\
&& \hspace{6.0 cm} + \left \{ \ln(r_3-1) \, \ln {(r_3-1)^3 \over (1-r_2)^2}
- \pi^2 \right \}  \, f(1) \,,
\label{master dispersion formula last}
\end{eqnarray}
where the ``plus" function is defined by
\begin{eqnarray}
\int_{r_2}^{r_3} d \rho \,\, [g(\rho)]_{+} \, f(\rho)
= \int_{r_2}^{r_3} d \rho \,\, g(\rho)\,  \left [ f(\rho) - f(1)  \right ] \,,
\end{eqnarray}
and the Heaviside step function $\theta(r_3-1)$ has been suppressed
on the right-hand sides of
(\ref{master dispersion formula first})-(\ref{master dispersion formula last}).

\section{Photon distribution amplitudes}
\label{appendix: Photon distribution amplitudes}

We summarize the operator-level definitions of the two-particle and three-particle photon distribution amplitudes
up to and including the twist-four accuracy, following the systematic classification detailed in \cite{Ball:2002ps},
and taking this opportunity correct several misprints in the previous expressions
displayed in Appendix B of \cite{Wang:2018wfj}.
\begin{eqnarray}
&& \langle \gamma(p) |\bar q(x) \, W_c(x, 0) \,\, \sigma_{\alpha \beta} \,\, q(0)| 0 \rangle  \nonumber \\
&& = - i \, g_{\rm em} \, Q_q \,  \langle \bar q q \rangle(\mu) \,
(p_{\beta} \, \epsilon_{\alpha}^{\ast} - p_{\alpha} \, \epsilon_{\beta}^{\ast}) \,
\int_0^1 \, d z \, e^{i \, z \, p \cdot x} \, \left [  \chi(\mu) \, \phi_{\gamma}(z, \mu)
+ {x^2 \over 16} \, \mathbb{A}(z, \mu) \right ]   \nonumber \\
&& \hspace{0.3 cm} - \, {i \over 2} \, g_{\rm em} \, Q_q \,  {\langle \bar q q \rangle(\mu) \over q \cdot x} \,
(x_{\beta} \, \epsilon_{\alpha}^{\ast} - x_{\alpha} \, \epsilon_{\beta}^{\ast}) \,
\int_0^1 \, d z \, e^{i \, z \, p \cdot x} \, h_{\gamma}(z, \mu) \,. \\
\nonumber \\
&& \langle \gamma(p) |\bar q(x) \, W_c(x, 0) \,\, \gamma_{\alpha} \,\, q(0)| 0 \rangle
=  g_{\rm em} \, Q_q \, f_{3 \gamma}(\mu) \, \epsilon_{\alpha}^{\ast} \,
\int_0^1 \, d z \, e^{i \, z \, p \cdot x} \, \psi^{(v)}(z, \mu)  \,. \\
\nonumber \\
&& \langle \gamma(p) |\bar q(x) \, W_c(x, 0) \,\, \gamma_{\alpha} \, \gamma_5 \,\, q(0)| 0 \rangle \nonumber \\
&& = {g_{\rm em} \over 4} \, Q_q \,  f_{3 \gamma}(\mu) \, \varepsilon_{\alpha \beta \rho \tau} \,
p^{\rho} \, x^{\tau} \,\epsilon^{\ast \, \beta} \, \int_0^1 \, d z \, e^{i \, z \, p \cdot x} \,
\, \psi^{(a)}(z, \mu)\,. \\
\nonumber \\
&& \langle \gamma(p) |\bar q(x) \, W_c(x, 0) \,\, g_s \, G_{\alpha \beta}(v \, x) \, \, q(0)| 0 \rangle \nonumber \\
&& =  - i \, g_{\rm em} \, Q_q \,  \langle \bar q q \rangle(\mu) \,
(p_{\beta} \, \epsilon_{\alpha}^{\ast} - p_{\alpha} \, \epsilon_{\beta}^{\ast}) \,
\int [{\cal D} \alpha_i] \, e^{i \, (\alpha_q +  \, v \, \alpha_g) \, p \cdot x} \, S(\alpha_i, \mu) \,. \\
\nonumber \\
&& \langle \gamma(p) |\bar q(x) \, W_c(x, 0) \,\, g_s \, \widetilde{G}_{\alpha \beta}(v \, x) \,
i \, \gamma_5 \,\,  q(0)| 0 \rangle \nonumber \\
&& =  i \, g_{\rm em} \, Q_q \,  \langle \bar q q \rangle(\mu) \,
(p_{\beta} \, \epsilon_{\alpha}^{\ast} - p_{\alpha} \, \epsilon_{\beta}^{\ast}) \,
\int [{\cal D} \alpha_i] \, e^{i \, (\alpha_q +  \, v \, \alpha_g) \, p \cdot x} \, \widetilde{S}(\alpha_i, \mu) \,. \\
\nonumber \\
&& \langle \gamma(p) |\bar q(x) \, W_c(x, 0) \,\, g_s \, \widetilde{G}_{\alpha \beta}(v \, x) \,
\gamma_{\rho} \, \gamma_5 \,\,  q(0)| 0 \rangle \nonumber \\
&& = - g_{\rm em} \, Q_q \, f_{3 \gamma}(\mu) \, p_{\rho} \,
(p_{\beta} \, \epsilon_{\alpha}^{\ast} - p_{\alpha} \, \epsilon_{\beta}^{\ast}) \,
\int [{\cal D} \alpha_i] \, e^{i \, (\alpha_q +  \, v \, \alpha_g) \, p \cdot x} \, A(\alpha_i, \mu) \,. \\
\nonumber \\
&& \langle \gamma(p) |\bar q(x) \, W_c(x, 0) \,\, g_s \, G_{\alpha \beta}(v \, x) \,
i \, \gamma_{\rho} \,\,  q(0)| 0 \rangle \nonumber \\
&& = - g_{\rm em} \, Q_q \, f_{3 \gamma}(\mu) \,  p_{\rho} \,
(p_{\beta} \, \epsilon_{\alpha}^{\ast} - p_{\alpha} \, \epsilon_{\beta}^{\ast}) \,
\int [{\cal D} \alpha_i] \, e^{i \, (\alpha_q +  \, v \, \alpha_g) \, p \cdot x} \, V(\alpha_i, \mu)  \,. \\
\nonumber \\
&& \langle \gamma(p) |\bar q(x) \, W_c(x, 0) \,\, g_{\rm em} \, Q_q \,  F_{\alpha \beta}(v \, x) \, \, q(0)| 0 \rangle \nonumber \\
&& =  - i \, g_{\rm em} \, Q_q \,  \langle \bar q q \rangle(\mu) \,
(p_{\beta} \, \epsilon_{\alpha}^{\ast} - p_{\alpha} \, \epsilon_{\beta}^{\ast}) \,
\int [{\cal D} \alpha_i] \, e^{i \, (\alpha_q +  \, v \, \alpha_g) \, p \cdot x} \, S_{\gamma}(\alpha_i, \mu) \,.  \\
\nonumber \\
&& \langle \gamma(p) |\bar q(x) \, W_c(x, 0) \,\, \sigma_{\rho \tau} \,\, g_s \, G_{\alpha \beta}(v \, x)
\,\,  q(0)| 0 \rangle \nonumber \\
&& =  \, g_{\rm em} \, Q_q \,\langle \bar q q \rangle(\mu) \,
\left [p_{\rho} \, \epsilon_{\alpha}^{\ast} \, g_{\tau \beta}^{\perp}
- p_{\tau} \, \epsilon_{\alpha}^{\ast} \, g_{\rho \beta}^{\perp} - (\alpha \leftrightarrow \beta) \right ]  \,
\int [{\cal D} \alpha_i] \, e^{i \, (\alpha_q +  \, v \, \alpha_g) \, p \cdot x} \, T_{1}(\alpha_i, \mu)  \nonumber \\
&& \hspace{0.4 cm} + \, g_{\rm em} \, Q_q \,\langle \bar q q \rangle(\mu) \,
\left [p_{\alpha} \, \epsilon_{\rho}^{\ast} \, g_{\tau \beta}^{\perp}
- p_{\beta} \, \epsilon_{\rho}^{\ast} \, g_{\tau \alpha}^{\perp} - (\rho \leftrightarrow \tau) \right ]  \,
\int [{\cal D} \alpha_i] \, e^{i \, (\alpha_q +  \, v \, \alpha_g) \, p \cdot x} \, T_{2}(\alpha_i, \mu) \nonumber \\
&& \hspace{0.4 cm} + \, g_{\rm em} \, Q_q \,\langle \bar q q \rangle(\mu) \,
\frac{(p_{\alpha} \, x_{\beta} - p_{\beta} \, x_{\alpha} ) (p_{\rho} \, \epsilon_{\tau}^{\ast} - p_{\tau} \,\epsilon_{\rho}^{\ast})}
{p \cdot x} \, \int [{\cal D} \alpha_i] \, e^{i \, (\alpha_q +  \, v \, \alpha_g) \, p \cdot x} \,
 T_{3}(\alpha_i, \mu) \nonumber \\
&& \hspace{0.4 cm}  + \, g_{\rm em} \, Q_q \,\langle \bar q q \rangle(\mu) \,
\frac{(p_{\rho} \, x_{\tau} - p_{\tau} \, x_{\rho} ) (p_{\alpha} \, \epsilon_{\beta}^{\ast} - p_{\beta} \,\epsilon_{\alpha}^{\ast})}
{p \cdot x} \, \int [{\cal D} \alpha_i] \, e^{i \, (\alpha_q +  \, v \, \alpha_g) \, p \cdot x} \,
 T_{4}(\alpha_i, \mu) \,. \\
 \nonumber \\
&& \langle \gamma(p) |\bar q(x) \, W_c(x, 0) \,\, \sigma_{\rho \tau} \,\, g_{\rm em} \, Q_q \, F_{\alpha \beta}(v \, x)
\,\,  q(0)| 0 \rangle \nonumber \\
&& =  \, g_{\rm em} \, Q_q \,\langle \bar q q \rangle(\mu) \,
\frac{(p_{\rho} \, x_{\tau} - p_{\tau} \, x_{\rho} ) (p_{\alpha} \, \epsilon_{\beta}^{\ast} - p_{\beta} \,\epsilon_{\alpha}^{\ast})}
{p \cdot x} \, \int [{\cal D} \alpha_i] \, e^{i \, (\alpha_q +  \, v \, \alpha_g) \, p \cdot x} \,
 T_{4}^{\gamma}(\alpha_i, \mu)  + ... \hspace{0.5 cm}
\end{eqnarray}
Here, we adopt the following convention for the dual gluon-field strength tensor
\begin{eqnarray}
\widetilde{G}_{\alpha \beta}= {1 \over 2} \,  \varepsilon_{\alpha \beta \rho \tau }  \, G^{\rho \tau} \,.
\end{eqnarray}


\end{document}